\documentclass{pasj01}

\Received{$\langle$reception date$\rangle$}
\Accepted{$\langle$acception date$\rangle$}
\Published{$\langle$publication date$\rangle$}

\usepackage{natbib,bm}
\usepackage[varg]{txfonts}  

\usepackage[pdftex]{graphicx,color}  

\definecolor{pinegreen}{cmyk}{0.92, 0.0, 0.59,0.25}

\newcommand{\pd}{\partial}

\newcommand{\paren}[1]{\left({#1}\right)} 

\newcommand{\eqref}[1]{(\ref{#1})}

\begin{document}

\title{Two saturated states of the vertical shear instability in protoplanetary disks with vertically varying cooling times}

\author{Yuya \textsc{Fukuhara}\altaffilmark{1}}
\author{Satoshi \textsc{Okuzumi}\altaffilmark{1}}
\author{Tomohiro \textsc{Ono}\altaffilmark{1}\altaffilmark{2}}

\altaffiltext{1}{Department of Earth and Planetary Sciences, Tokyo Institute of Technology, Meguro, Tokyo 152-8551, Japan}
\altaffiltext{2}{School of Natural Sciences, Institute for Advanced Study, Princeton, NJ 08544, USA}
\email{fukuhara.y.ab@m.titech.ac.jp}


\maketitle

\begin{abstract}
    Turbulence in protoplanetary disks plays an important role in dust evolution and planetesimal formation.
    The vertical shear instability (VSI) is one of the candidate hydrodynamic mechanisms that can generate turbulence in the outer disk regions.
    The VSI requires rapid gas cooling in addition to vertical shear.
    A linear stability analysis suggests that the VSI may not operate around the midplane where gas cooling is inefficient.
    In this study, we investigate the nonlinear outcome of the VSI in disks with a linearly VSI-stable midplane region.
    We perform two-dimensional global hydrodynamical simulations of an axisymmetric disk with vertically varying cooling times. 
    The vertical cooling time profile determines the thicknesses of the linearly VSI-stable midplane layer and unstable layers above and below the midplane.
    We find that the thickness of the midplane stable layer determines the vertical structure of VSI-driven turbulence in the nonlinear saturated state.
    We identify two types of final saturated state: (1) T states characterized by vertical turbulent motion penetrating into the VSI-stable midplane layer and (2) pT states characterized by turbulent motion confined in the unstable layers.
    The pT states are realized when the midplane VSI-stable layer is thicker than two gas scale heights. 
    We also find that the VSI-driven turbulence is largely suppressed at all heights when the VSI-unstable region lying above and below the midplane is thinner than two gas scale heights.
    We present empirical formulas that predict the strength of VSI-driven turbulence as a function of the thicknesses of the unstable and stable layers.
    These formulas will be useful for studying how VSI-driven turbulence and dust grains controlling the disk cooling efficiency evolve simultaneously. 
\end{abstract}


\section{Introduction}\label{sec:intro}
The initial stage of planet formation is the formation of kilometer-sized planetesimals from micron-sized dust grains in protoplanetary disks (see \citealt{Johansen+2014} and \citealt{DrazkowskaBitsch+:2022qi} for recent reviews).
Micrometer-sized grains initially grow into larger aggregates through coagulation. 
Sufficiently large aggregates settle onto the disk midplane and form a dense dust layer \citep{Weidenschilling:1980xl,NakagawaNakazawa+:1981wj,DullemondDominik:2005vy,TanakaHimeno+:2005wc}.
Large aggregates can also concentrate radially toward local maxima of the disk gas pressure (e.g., \citealt{Whipple:1972vv,KretkeLin:2007mn,PinillaBirnstiel+:2012vz}).
These dust concentration processes may lead to planetesimal formation through the streaming instability \citep{Youdin:2005aa,JohansenYoudin2007,Johansen:2009aa,Carrera:2015aa,Yang:2017aa} and gravitational instabilities \citep{Goldreich:1973aa,Youdin:2011aa,Takahashi:2014wi,Tominaga:2018th,Tominaga:2019uu,Tominaga:2020wn,Pierens:2021aa}.
If the aggregates are sticky enough, they may form planetesimals directly by successive coagulation (e.g. \citealt{Okuzumi+2012,Windmark:2012aa,Kataoka:2013aa}).

In these planetesimal formation processes, gas disk turbulence can play many important roles, including both positive and negative ones.
Large long-lived vortices created by turbulence may lead to dust concentration and subsequent planetesimal formation through gravitational collapse (e.g., \citealt{BargeSommeria:1995qd,RaettigLyra+:2021sb,LehmannLin:2022nr}).
On the other hand, strong turbulence induces a large impact velocity between dust particles (e.g., \citealt{OrmelCuzzi2007}) that may inhibit planetesimal formation through coagulation (e.g., \citealt{Brauer:2008aa,Okuzumi:2012aa}).
Turbulence can also diffuse the midplane dust layer and prevent the onset of gravitational and streaming instabilities (e.g., \citealt{Dubrulle+1995,UmurhanEstrada+:2020yi,ChenLin:2020kh}).
Therefore, the strength of turbulence would determine how planetesimals form.

Recent radio interferometric observations with the Atacama Large Millimeter--submillimeter Array (ALMA) and other interferometers have provided important constraints on turbulent intensity in the outer region of protoplanetary disks.
Molecular line emission observations point to weak or no turbulence in the upper layers \citep{HughesWilner+:2011ed,Flaherty:2015aa,Flaherty:2017aa,Flaherty:2018aa,Flaherty:2020aa} and regions within one scale height above and below the midplane \citep{GuilloteauDutrey+:2012dd,TeagueGuilloteau+:2016uj,TeagueHenning+:2018wx} in the outer disk region (for a recent review, see \citealt{PinteTeague+:2022om}).
Strong nonthermal gas motion is detected in a disk around DM Tau \citep{Flaherty:2020aa}.
Observations of dust rings and gaps in massive and large disks (e.g., \citealt{ALMA+2014,Andrews+2018,Long:2018aa,van-der-Marel:2019aa}) can also be used to indirectly constrain the turbulence intensity because the morphology of the rings and gaps can be affected by turbulent dust diffusion. 
The well-defined morphology of the dust gaps in the disks around HL Tau \citep{Pinte:2016aa} and Oph163131 \citep{VillenaveStapelfeldt+:2022pp} suggests a significantly low level of turbulence.
On the other hand, one of the two major dust rings in the HD 163296 disk shows a high degree of vertical dust diffusion \citep{DoiKataoka:2021oz}, potentially implying strong turbulent diffusion in that location.
Taken together, the observations so far suggest that the outer disk regions are mostly laminar but can become strongly turbulent under some circumstances. 

To better understand when and where strong turbulence emerges, it is essential to study what mechanisms drive disk turbulence.
Regarding the outer disk region, it was previously believed that the magnetorotational instability (MRI; \citealt{BalbusHawley1991}) is the most viable mechanism for driving turbulence (e.g., \citealt{SanoMiyama+:2000fo}). 
However, recent theoretical studies have shown that the viability of the MRI in the outer disk is limited by ambipolar diffusion \citep{Simon+2013a,Simon+2013b,Bai2015,BethuneLesur+:2017aa,Riols:2018aa,CuiBai:2021aa}.
Instead, recent studies point out the importance of purely hydrodynamic instabilities (for reviews, see \citealt{LyraUmurhan2019} and \citealt{LesurErcolano+:2022kp}), in particular the vertical shear instability \citep{UrpinBrandenburg1998,ArltUrpin2004,NelsonGresselUmurhan2013,LinYoudin2015}.
This is a hydrodyamical instability similar to the Goldreich--Schubert--Fricke instability in differentially rotating stars \citep{GS67,Fricke:1968aa}.
This instability requires a vertical gradient in the gas orbital velocity and rapid cooling of disk gas \citep{Urpin2003,NelsonGresselUmurhan2013,LinYoudin2015,MangerPfeil+:2021cm}.
Turbulence driven by the VSI has a predominant vertical motion (e.g., \citealt{NelsonGresselUmurhan2013,StollKley2014}) that can prevent dust settling  \citep{StollKley:2016vp,FlockNelson+2017,Flock:2020aa}. 
Because the VSI requires rapid gas cooling, it is most active in outer disk regions with low optical depths \citep{Malygin+2017,PfeilKlahr2019,FukuharaOkuzumi+:2021ca}.
In the outer regions, the VSI can take over the MRI \citep{Cui:2020aa,CuiBai:2022aa}.

Despite its importance, it is yet to be fully understand in what conditions the VSI produces turbulence around the midplane, where dust evolution and planetesimal formation mainly take place.
According to local linear analysis, the VSI tends to be suppressed in a midplane region with a high optical depth and hence a low cooling rate \citep{Malygin+2017,PfeilKlahr2019}.
However, \cite{PfeilKlahr:2021nr} recently showed that such an optically thick midplane region can become turbulent if the VSI operates above the region. 
They performed hydrodynamical simulations using a vertically varying cooling rate profile and found that a vertical gas motion generated by the VSI crosses the midplane in the final saturated state.
Yet, it remains unclear how strongly their results depend on the assumed vertical profile of the gas cooling rate. 
In general, the disk cooling profile is determined by the size and spatial distribution of dust grains that dominate the disk opacity \citep{Malygin+2017}. 
Therefore, the cooling profile can change significantly as the grains  grow and/or settle toward the midplane \citep{BarrancoPei+:2018kc,FukuharaOkuzumi+:2021ca}.
Very recently, \citet{DullemondZiampras+:2022aa} also showed that depletion of sub-micrometer-sized dust grains due to coagulation increases the gas cooling time and makes the disk VSI-stable.
These aspects are particularly important in the context of planetesimal formation, where dust evolution is inevitable.

In this paper, we investigate systematically how the saturated state of VSI-driven turbulence depends on the vertical profile of the disk cooling rate.
We perform global two-dimensional (2D) hydrodynamical simulations of an axisymmetric protoplanetary disk with a parameterized vertical profile of the disk gas cooling timescale.
We show that, depending on the thickness of the VSI-stable midplane region, a vertical gas motion generated by the VSI above the midplane either does or does not penetrate into the midplane.
We also provide empirical formulas that predict the saturated level of VSI-driven turbulence from a given vertical profile of the gas cooling timescale.
Such empirical relations will be useful to study how the saturated state of VSI-driven turbulence varies with long-term evolution of the background disk. 

This paper is organized as follows. 
In section \ref{sec:method}, we describe our numerical hydrodynamic simulation setup, our disk cooling (thermal relaxation) model, and a simulation analysis method.
We then present the main results in section \ref{sec:results} and discuss the implications of our study in section \ref{sec:discussion}.
Section \ref{sec:summary} presents a summary.

\section{Numerical method}\label{sec:method}

We perform global 2D hydrodynamical simulations of an axisymmetric protoplanetary disk in spherical polar coordinates $(r,~\theta,~\phi)$.
We also use the cylindrical radius $R = r\sin{\theta}$ and vertical coordinate $z = r\cos{\theta}$.
The solved continuity, motion, and energy equations of hydrodynamics are 
\begin{equation}\label{eq:continuity}
    \frac{\pd \rho_{\rm g}}{\pd t}+ \nabla \cdot \paren{\rho_{\rm g}\bm v} = 0,
\end{equation}
\begin{equation}\label{eq:EOM}
    \frac{\pd \rho_{\rm g} {\bm v}}{\pd t} + \nabla\cdot\left(\rho_{\rm g}{\bm v}{\bm v} \right) = -\nabla P-\rho_{\rm g}\nabla \Phi,
\end{equation}
\begin{equation}\label{eq:energy_eq}
    \frac{\pd E}{\pd t}+\nabla \cdot \left[{\bm v}\paren{E+P} \right] = -\rho_{\rm g}{\bm v}\cdot\nabla \Phi -\frac{U\paren{\rho_{\rm g},~T}-U\paren{\rho_{\rm g},~T_0}}{t_{\rm relax}},
\end{equation}
where $\rho_{\rm g}$ is the gas density, $\bm v$ is the gas velocity, $P$ is the gas pressure, $\Phi$ is the gravitational potential, $E$ is the total energy per unit volume, $U$ is the internal energy per unit volume, $T$ is the temperature, and $T_0$ is the temperature before the cooling. 
The gas velocity has three components for the radial, meridional, and azimuthal in spherical polar coordinates ${\bm v} = (v_r,~v_\theta,~v_\phi)$.
The gravitational potential is given by $\Phi = -GM/R$, where $G$ is the gravitational constant and $M$ is the mass of the central star.
The total energy per unit volume $E$ and internal energy per unit volume $U$ are
\begin{equation}
    E = U\paren{\rho_{\rm g},~T} + \frac{1}{2}\rho_{\rm g} \mbox{\bm $v$}^2,
\end{equation}
and
\begin{equation}
    U\paren{\rho_{\rm g},~T} = \frac{P\paren{\rho_{\rm g},~T}}{\gamma-1},
\end{equation}
respectively, where $\gamma$ is the heat capacity ratio and is taken to be 1.4.
In equation \eqref{eq:energy_eq}, $t_{\rm relax}$ is the thermal relaxation (cooling) timescale.
In this study, the thermal relaxation timescale is an important parameter that determines the strength of VSI-driven turbulence and is set with a spatial distribution.
We describe more details of $t_{\rm relax}$ model in section \ref{subsec:beta}.

To solve the hydrodynamical equations (equations \eqref{eq:continuity}--\eqref{eq:energy_eq}), we use the Athena++ code \citep{Stone:2020aa} with the combination of the Harten--Lax--Van Leer (HLLC) approximate Riemann solver \citep{MignoneBodo:2005bv}, the reconstruction scheme of a second-order piecewise linear method \citep{van-Leer:1974xu}, and second-order Runge--Kutta time integrator.
The Courant--Friedrichs--Lewy (CFL) number is set to 0.3.

    \subsection{Simulation setup}\label{subsec:setup}
    
    \begin{table}
    \tbl{Parameters choices of global 2D hydrodynamical simulations.}{%
    \begin{tabular}{lcc}
    \hline
    Parameter & Symbol & Value \\
    \hline
    Reference gas density & $\rho_0$ & 1.0 \\
    Reference radius & $R_0$ & 1.0 \\ 
    Radial power–law index for the gas density & $p$ & $-2.25$ \\
    Reference sound speed & $c_0$ & 0.05 \\
    Radial power–law index for the temperature & $q$ & $-0.5$ \\
    Reference gas scale height & $H_0$ & 0.05 \\
    \hline
    \end{tabular}}\label{t:setup}
    \end{table}

    We assume that the disk is initially in vertical hydrostatic equilibrium and give the initial gas density profile as
    \begin{equation}\label{eq:initial_gas_density}
        \rho_{\rm g} = \rho_0 \paren{\frac{R}{R_0}}^p \exp{\left(-\frac{z^2}{2H_{\rm g}^2} \right)},
    \end{equation}
    where $\rho_0$ is the reference gas density, $R_0$ is the reference radius, and $p$ is the radial power-law index for the gas density. 
    The sound speed is given by
    \begin{equation}
        c_{\rm s} = c_0\paren{\frac{R}{R_0}}^{q/2},
    \end{equation}
    where $c_0$ is the reference sound speed and $q$ is the radial power-law index of the temperature.
    The disk scale height is 
    \begin{equation}
        H_{\rm g} = \frac{c_{\rm s}}{\Omega_{\rm K}} = H_0\paren{\frac{R}{R_0}}^{(q+3)/2},
    \end{equation}
    where $\Omega_{\rm K} = \sqrt{GM/R^3}$ is the Keplerian frequency and $H_0 = c_0/\Omega_{\rm K}$ is the reference gas scale height.
    Our parameter choices are summarized in table \ref{t:setup}.
    
    The initial velocities are set to $v_r = v_\theta = 0$ and $v_\phi = R\Omega\paren{R,~z}$, where $\Omega\paren{R,~z}$ is the gas angular velocity. For $H_{\rm g} \ll R$, which is the case for the disk model in our simulations, $\Omega\paren{R,~z}$ can be approximated as \citep{TakeuchiLin2002}
    \begin{equation}\label{eq:gas_angular_velocity}
       \Omega\paren{R,~z} = \Omega_{\rm K} \left[ 1+\frac{1}{2}\paren{\frac{H_{\rm g}}{R}}^2 \paren{p+q+\frac{q}{2}\frac{z^2}{H_{\rm g}^2}} \right].
    \end{equation}
    From equation \eqref{eq:gas_angular_velocity}, the vertical shear of the gas rotation velocity $\pd(R\Omega)/\pd z$ is given by
    \begin{equation}\label{eq:vertical_shear}
        \frac{\pd (R\Omega)}{\pd z} = \frac{q}{2}\frac{z}{R}\Omega_{\rm K}.
    \end{equation}
    This physical quantity is the driving force behind the dynamics of the VSI and characterizes the strength of VSI-driven turbulence.
    We also add small cellwise random velocities with an amplitude of $10^{-5} = 2 \times 10^{-4}c_0$ to the initial velocity field.
    
    At all computational boundaries, we fix both density and pressure to the initial values. 
    For the velocity components normal to the boundaries, we apply the outflow boundary conditions preventing inflow at the inner and outer radial boundaries and reflecting boundary conditions at the upper and lower meridional boundaries.
    
    Grid cells have logarithmically and linearly uniform spacings in the radial and meridional directions, respectively.
    The radial and meridional domains cover $0.5 \leq r \leq 2.5$ and $\pi/2 - 5H_0/R_0 \leq \theta \leq \pi/2 + 5H_0/R_0$.
    Because a resolution of 100 cells or more per scale height is necessary to resolve VSI-driven turbulence \citep{Flores-Rivera:2020ab}, we adopt a resolution of 128 cells per scale height in both radial and vertical directions.
    Therefore, our simulations use 4160$\times$1280 grid cells\footnote{The highest resolution simulation in \citet{Flores-Rivera:2020ab} has 4096 cells in the radial domain extending from 0.5 to 5.0 and thereby has $\sim 100$ cells per gas scale height in the radial direction.
    }.
    
    We adopt the code units $M = G = R_0 = 1$.
    In this unit system, the orbital period at $R=R_0$ is $P_{\rm in} = 2\pi$.
    
    \subsection{Cooling model}\label{subsec:beta}
    We parameterize the thermal relaxation timescale $t_{\rm relax}$ in terms of the dimensionless cooling (thermal relaxation) time
    \begin{equation}\label{eq:beta}
        \beta = t_{\rm relax}\Omega_{\rm K}.
    \end{equation}
    We use the terms ``thermal relaxation" and ``cooling" interchangeably throughout this study.
    
    The cooling time varies vertically depending on the optical depth from the surface \citep{Malygin+2017}.
    In a region that is optically thin to its own thermal emission, the cooling time is primarily determined by the collision time for gas molecules and grains, and therefore scales inversely with the density and cross section of dust grains. 
    In an optically thick region, the diffusion approximation applies, and the cooling time scales linearly with the opacity.
    In both cases, the exact vertical dependence of the cooling time would depend on the size and vertical distribution of the grains. 
    To study the $\beta$ dependence of our simulations systematically, we parameterize $\beta$ as
    \begin{equation}\label{eq:beta_model}
        \beta (R,z) = \left[\beta_0\exp{\paren{\frac{z^2}{a^2H_{\rm g}^2}}} + \beta_1\exp{\paren{-\frac{z^2}{b^2H_{\rm g}^2}}} \right]\left(\frac{R}{R_0}\right)^{(q+1)/2},
    \end{equation}
    where $\beta_0$, $a$, $\beta_1$, and $b$ are dimensionless numbers that characterize the $\beta$ profile. 
    The first and second terms in the square bracket mimic the increase and decrease of $\beta$ with $|z|$ in the optically thin (high-$|z|$) and thick (low-$|z|$) regions, respectively, with $aH_{\rm g}$ and $bH_{\rm g}$ representing the vertical length scales over which $\beta$ varies.
    The radial dependence of $\beta$ is chosen to be the same as one of $H_{\rm g}/R$ to investigate the dependence of the radial constant thicknesses of VSI-unstable and stable layers determined by the critical cooling time of the VSI (see also section \ref{subsec:stable_unstable_layer}).
    In reality, $\beta$ can have a different radial dependence from this (e.g., \citealt{Malygin+2017}).
    In the simulations, we take $\beta$ not to exceed $10$ to avoid excessively large values of $\beta$ at high-$|z|$.
    
    In this study, we treat $a$ and $b$ as free parameters.
    In reality, these values can vary with dust growth and settling.
    The value of $a$ can increase or decrease depending on dust growth and settling.
    This is because the optically thick area around the midplane widens for a certain amount of dust concentration due to the presence of regions that change from a high to low optical depth, and narrows beyond that point.
    On the other hand, dust growth and settling cause dust depletion and increase cooling time at high altitudes \citep{BarrancoPei+:2018kc,FukuharaOkuzumi+:2021ca}, which corresponds to a decrease in $b$.
    
    \begin{figure}
        \begin{center}
        \includegraphics[width=80mm,bb = 0 0 566 561]{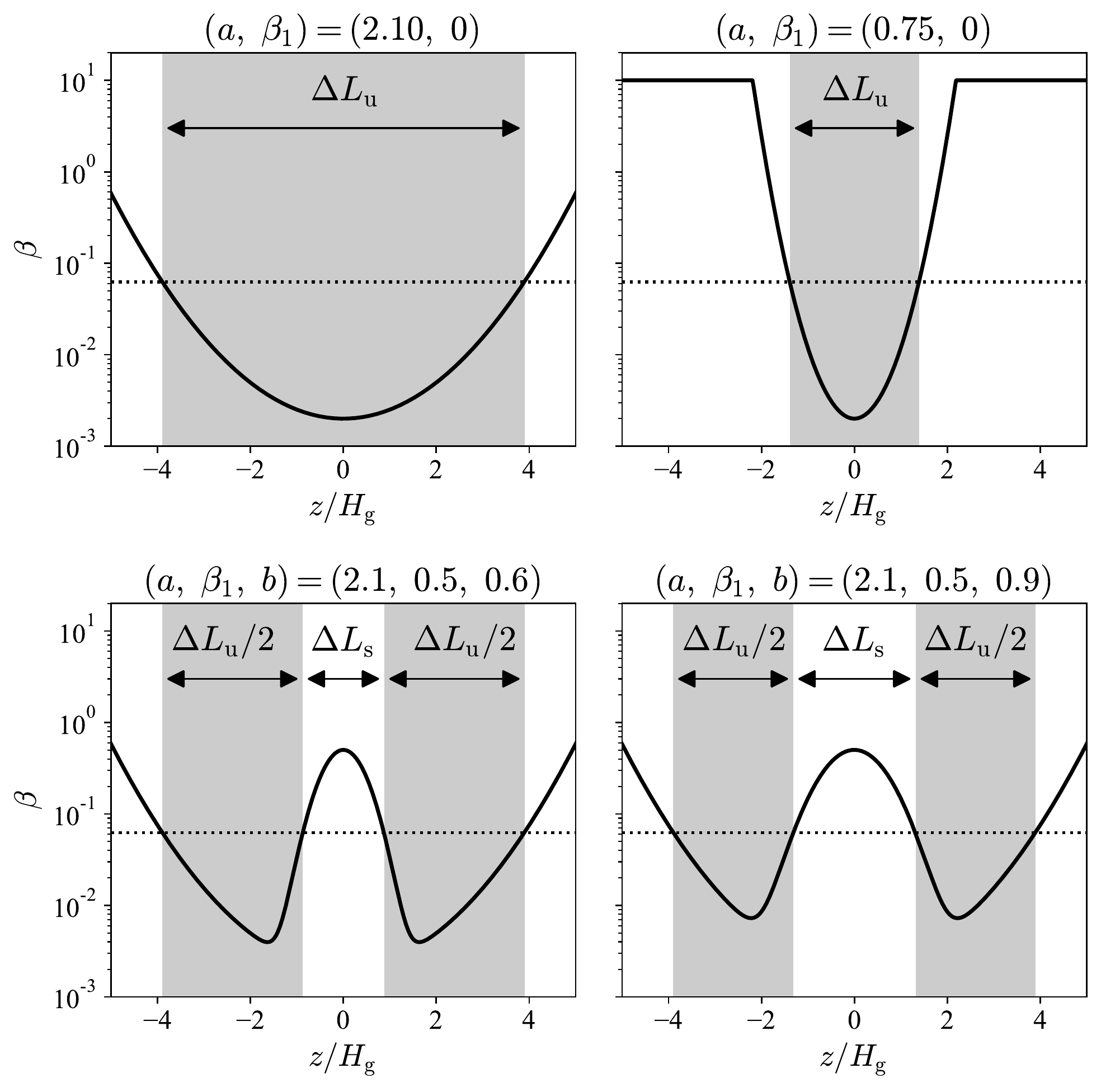}
        \end{center}
        \caption{Vertical $\beta$ profile for runs of $(a,~\beta_1) = (2.10,~0)$ and $(0.75,~0)$ (upper panels) and $(a,~\beta_1,~b) = (2.1,~0.5,~0.6)$ and $(2.1,~0.5,~0.9)$ (lower panels) at $R=1.0$. The gray regions are marked as ``unstable layer" and other white regions are marked as ``stable layer", which are determined by applying the linear criterion (equation \eqref{eq:global_criterion}). The dotted lines represent $\beta = \beta_{\rm gc}$. We refer to the thicknesses of the unstable layer (all panels) and stable midplane layer (only lower panels) as $\Delta L_{\rm u}$ and $\Delta L_{\rm s}$, respectively.}
        \label{fig:beta_zpro}
    \end{figure}
    
    We perform 46 simulations with different values of $a$, $\beta_1$, and $b$ with $\beta_0 = 2\times 10^{-3}$ as summarized in table \ref{t:beta} of appendix \ref{appendix:parameter_sets}.
    We also perform one simulation with the locally isothermal equation of state.
    We take $\beta$ to be constant in time.
    Figure \ref{fig:beta_zpro} illustrates the vertical profile of $\beta$ for four runs.    
    The $\beta$ profiles for all our runs are shown in figure \ref{fig:beta_model} in appendix \ref{appendix:parameter_sets}.
    
    \subsection{Linearly unstable and stable layers}\label{subsec:stable_unstable_layer}
    
    The prescribed vertical cooling rate profile determines where the linear growth of the VSI occurs.
    Following \citet{Malygin+2017} and  \citet{FukuharaOkuzumi+:2021ca} expected linear VSI growth in regions that fulfill the linear instability criterion, we use
    \begin{equation}\label{eq:global_criterion}
        \beta \lesssim \beta_{\rm gc},
    \end{equation}
    where 
    \begin{equation}\label{eq:global_cooling_time}
        \beta_{\rm gc} = \frac{|q|}{\gamma-1}\frac{H_{\rm g}}{R}
    \end{equation}
    is the dimensionless critical cooling time \citep{LinYoudin2015}.
    In this study, we refer to such regions the linearly unstable layers\footnote{\citet{FukuharaOkuzumi+:2021ca} called these the VSI zones.}.    
    \cite{LinYoudin2015} originally proposed equation \eqref{eq:global_criterion} as a criterion for the vertically global instability (see also appendix \ref{appendix:local_criterion}). 
    However, we find that this criterion well predicts where the VSI starts to grow in our simulations with vertically varying cooling times (see section \ref{sec:results}).
    
    All our simulations have linearly unstable layers at some heights (see  figure~\ref{fig:beta_zpro} for illustrative examples). 
    Runs with $\beta_1 = 0$ have a cooling time monotonically decreasing toward the midplane, yielding a single unstable layer at $|z| < z_{\rm u}$, where $z_{\rm u}$ is the height of the unstable layer's upper bundary.
    Runs with $\beta_1 > 0$ have two linearly unstable layers sandwiching a midplane region where equation~\eqref{eq:global_criterion} breaks down.
    We call this midplane region the linearly stable layer and denote the height of its boundary by $z_{\rm s}$. 
    When the linearly stable layer is absent, we set $z_{\rm s} = 0$.  
    
    We define the thicknesses of the linearly stable and unstable layers as $\Delta L_{\rm s} = 2z_{\rm s}$ and $\Delta L_{\rm u} = 2z_{\rm u} - \Delta L_{\rm s}$. When $z_{\rm s} > 0$, $\Delta L_{\rm u}$ accounts for the thicknesses of the two separated unstable layers lying at $z < 0$ and $z > 0$ (see figure~\ref{fig:beta_zpro}).
    In general, $\Delta L_{\rm u}$ decreases with decreasing $a$, and $\Delta L_{\rm s}$ increases with increasing $b$.
    The values of $\Delta L_{\rm u}$ and $\Delta L_{\rm s}$ for all runs are summarized in table \ref{t:beta} of appendix \ref{appendix:parameter_sets}.
    As we show in section \ref{sec:results}, $\Delta L_{\rm u}$ and $\Delta L_{\rm s}$ are key quantities that dictate the saturated state of VSI-driven turbulence.    
    
    \subsection{Turbulence diagnostics}\label{subsec:simulation_analysis}
    We quantify the strength of VSI-driven turbulence using the time average of the squared vertical velocity $\langle v_z^2\rangle$, where $v_z = v_r\cos{\theta}-v_\theta\sin{\theta}$ is the vertical velocity. 
    The bracket $\langle\cdots\rangle$ denotes the time average.
    In our simulations, the time averaging is performed after the system relaxes into a quasi-steady state.
    The mean squared vertical velocity is related to the vertical diffusion coefficient for gas and small dust particles \citep{FromangPapaloizou:2006rz}.
    We discuss this in more detail in section \ref{subsec:estimate_vartical_diffusion}.
    
    Turbulence also transports the disk's radial angular momentum.
    The efficiency of angular momentum transport is measured by the Reynolds stress $\langle\rho_{\rm g}\delta v_r\delta v_\phi\rangle$, where $\delta v_r = v_r -\langle v_r \rangle$ and $\delta v_\phi = v_\phi -\langle v_\phi \rangle$ are the dispersions of the radial and azimuthal velocity, respectively.
    
    In this study, we use the dimensionless Reynolds stress $\alpha_{r\phi}$ defined by
    \begin{equation}
        \alpha_{r\phi} = \frac{\langle\rho_{\rm g}\delta v_r\delta v_\phi\rangle}{\langle P \rangle}.
    \end{equation}
    We compute the spatial distribution of $\alpha_{r\phi}$ from our simulation results to estimate any possible radial angular momentum transport caused by the VSI-driven turbulent motions.
    Furthermore, to see a global effect of turbulence, we calculate the vertical average of $v_z^2$ and $\alpha_{r\phi}$ from $z = -z_{\rm u}$ to $z = z_{\rm u}$, which is denoted by overbars.
    
    The time that it takes for the system to reach a quasi-steady state differs from one run to another (see figure \ref{fig:vz_colormap_tz}), which may depend on the unstable layer thickness.
    Therefore, we stop a run at 400 orbits if the quasi-steady state has already been reached by 250 orbits; otherwise, we continue the run until 800 orbits.
    The stopping times for all runs are summarized in table \ref{t:beta} of appendix \ref{appendix:parameter_sets}.
    For simulations with shorter and longer runtimes, time averaging is performed over periods of 250--400 and 650--800 orbits, respectively.
    
\section{Results}\label{sec:results}
In this section, we present our simulation results to study how the vertical profile of the cooling time affects VSI-driven turbulence in protoplanetary disks.
We begin by defining two saturated states of VSI-driven turbulence and then analyze how the saturated state depends on the thicknesses of the unstable and stable layers in section \ref{subsec:overview}.
We construct empirical formulas of $\langle v_z^2\rangle|_{\rm mid}$, $\langle \overline{v_z^2}\rangle$, and $\overline{\alpha_{r\phi}}$ as a function of the thicknesses of the unstable and midplane stable layers, i.e., $\Delta L_{\rm s}$ and $\Delta L_{\rm u}$, in section \ref{subsec:fitting}.
    
    \subsection{Two saturated states of VSI-driven turbulence }\label{subsec:overview}
    
    \begin{figure*}
        \begin{center}
        \includegraphics[width=\hsize,bb = 0 0 841 323]{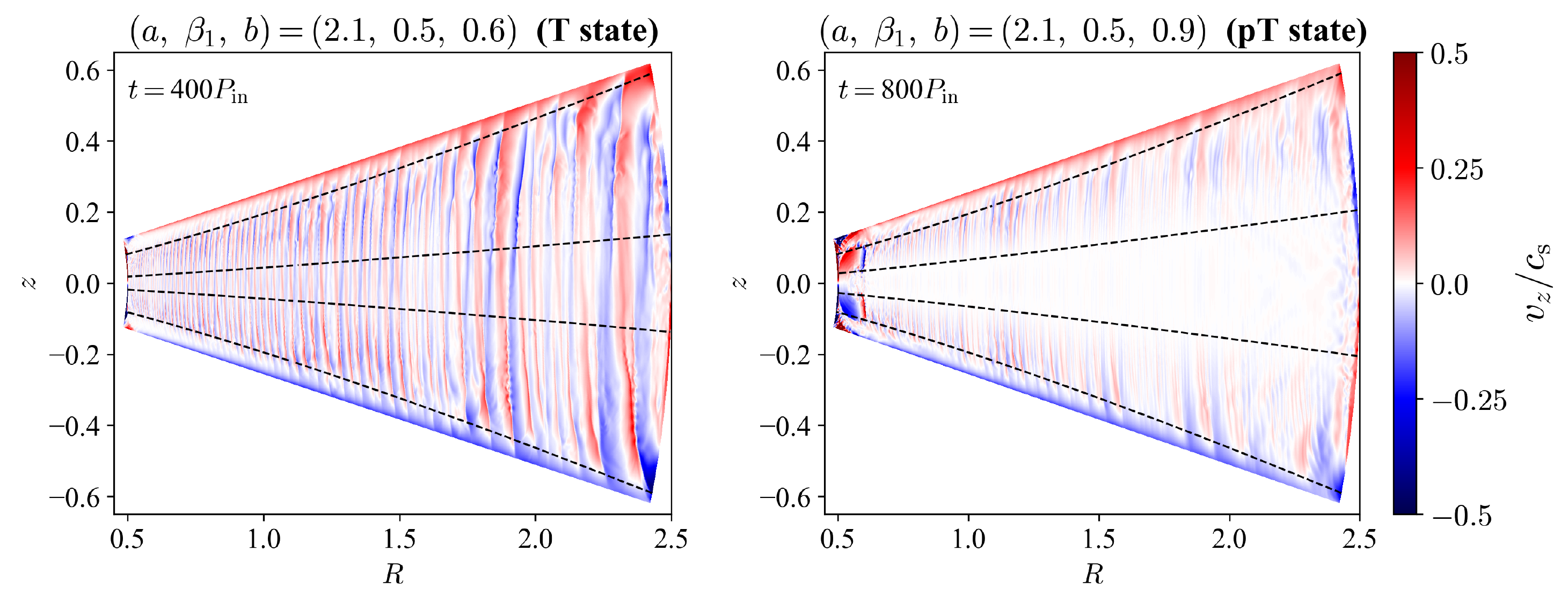}
        \end{center}
        \caption{Vertical velocity $v_z/c_{\rm s}$ for runs with $(a,~\beta_1,~b) = (2.1,~0.5,~0.6)$ (left panel) and $(a,~\beta_1,~b) = (2.1,~0.5,~0.9)$ (right panel), which relaxes into the T state and pT state, respectively, as a function of $R$ and $z$ at the end of the simulations. The dashed lines indicate the unstable layer's upper boundaries at $z = \pm z_{\rm u}$ (uppermost and lowermost lines) and stable midplane layer's upper boundaries at $z = \pm z_{\rm s}$ (two middle lines).}
        \label{fig:vz_colormap_Rz}
    \end{figure*}
    
    We find that the width of the linearly VSI-stable layer at the midplane determines the vertical structure of VSI-driven turbulence in a steady state.
    Specifically, we identify two possible saturated states of turbulence.
    In the first class of saturated states, which we call {\it the T (turbulent) states}, the vertical gas motion generated in the linearly VSI-unstable layers penetrates into the linearly VSI-stable midplane. 
    This state was already seen in the simulations by \citet{PfeilKlahr:2021nr}.
    In the second class, which we call {\it the pT (partially turbulent) states}, the vertical gas motion is well confined in the VSI-unstable layers, leaving the VSI-stable midplane layer only weakly turbulent. 
    
    Figure \ref{fig:vz_colormap_Rz} illustrates the two saturated states.
    Here, we present the two-dimensional maps of the vertical velocity $v_z$, normalized by the sound speed $c_{\rm s}$, at the end of the simulations for $(a,~\beta_1,~b) = (2.1,~0.5,~0.6)$ and $(2.1,~0.5,~0.9)$.
    The run with $(a,~\beta_1,~b) = (2.1,~0.5,~0.6)$ relaxes into the T state with a vertically uniform gas motion through $|z| < z_{\rm u} = 3.9 H_{\rm g}$.
    This state is not expected from the linear stability analysis, which predicts that the midplane region of $|z| < z_{\rm s} = 0.9H_{\rm g}$ is linearly VSI-stable.
    The run with $(a,~\beta_1,~b) = (2.1,~0.5,~0.9)$ relaxes into the pT state with little vertical gas motion inside the linearly VSI-stable layers ($z_{\rm s} = 1.3H_{\rm g}$ for this case).
    
    Our simulations do not clearly reproduce a finer spatial profile of vertical velocity caused by the secondary parametric instability, which can be related to the nonlinear saturation process of the VSI.
    This is because our simulation resolution is insufficient to resolve the parametric instability that may require $\sim 300$ cells per gas scale height in the radial direction \citep{CuiLatter:2022aa}, while it is sufficient to resolve the dominant VSI modes \citep{Flores-Rivera:2020ab}.
    
    \begin{figure*}
        \begin{center}
        \includegraphics[width=\hsize,bb = 0 0 843 329]{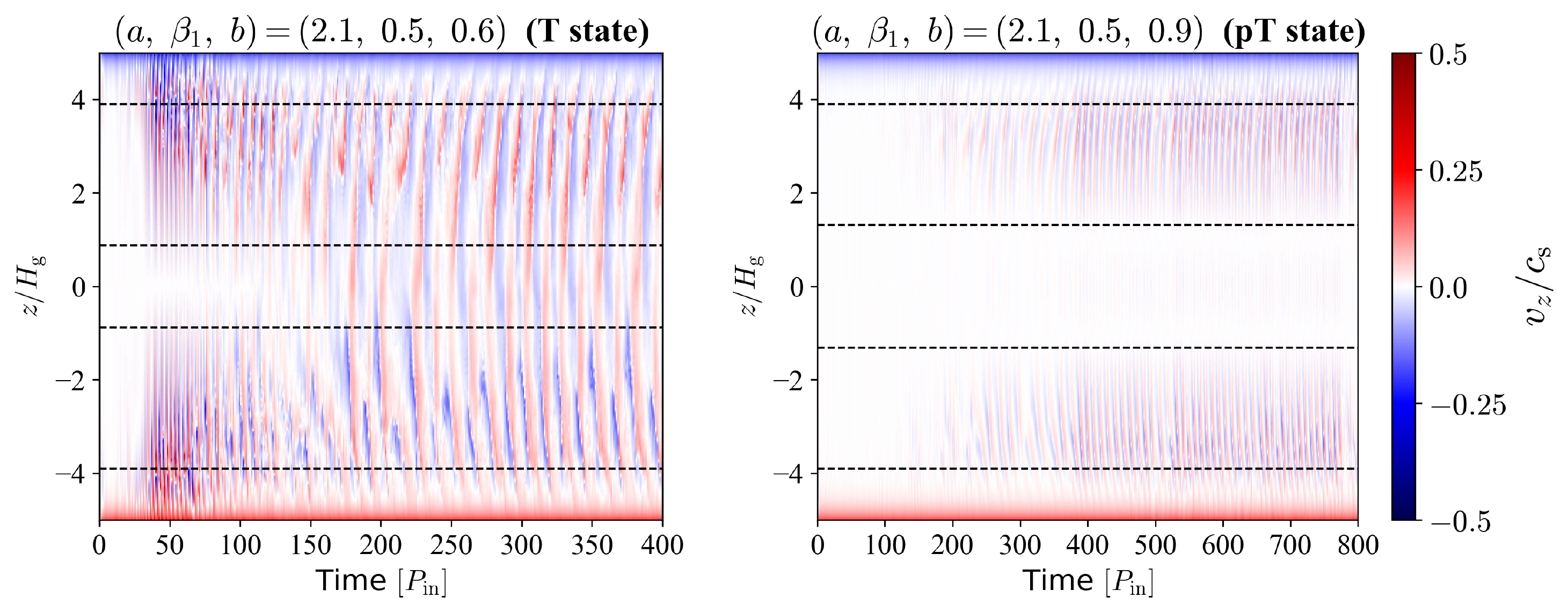}
        \end{center}
        \caption{Vertical velocity $v_z/c_{\rm s}$ for runs with $(a,~\beta_1,~b) = (2.1,~0.5,~0.6)$ (T state; left panel) and $(a,~\beta_1,~b) = (2.1,~0.5,~0.9)$ (pT state; right panel) as a function of time and $z/H_{\rm g}$ at $R=1.0$. The dashed lines indicate $z = \pm z_{\rm u}$ (top and bottom lines) and $z = \pm z_{\rm s}$ (two middle lines).}
        \label{fig:vz_colormap_tz}
    \end{figure*}
    
    To see how the final saturated states are reached, we plot in figure \ref{fig:vz_colormap_tz} the vertical profiles of $v_z$ at $R=1.0$ for the two cases displayed in figure \ref{fig:vz_colormap_Rz}.
    In both cases, the vertical gas motion starts to develop near the unstable layer's upper boundary at $|z|= z_{\rm u}$, where the vertical shear is strong.
    For $(a,~\beta_1,~b) = (2.1,~0.5,~0.6)$ (T state), the vertical flows developed in the VSI-unstable layers overshoot the stable midplane layer and eventually form a unified flow (see also \citet{PfeilKlahr:2021nr}).
    For $(a,~\beta_1,~b) = (2.1,~0.5,~0.9)$ (pT state), one can see that the gas vertical motion is well confined in the unstable layers. 
    The vertical profile of $\langle v_z^2\rangle$ is steady over the $t = 400$--$800$ orbits, suggesting that our simulation captures the final saturated state.
    Figures \ref{fig:vz2_TE} in appendix \ref{appendix:comparison} presents the time evolution of turbulence diagnostics for the two runs presented here.
    
    \begin{figure}
        \begin{center}
        \includegraphics[width=80mm,bb = 0 0 300 284]{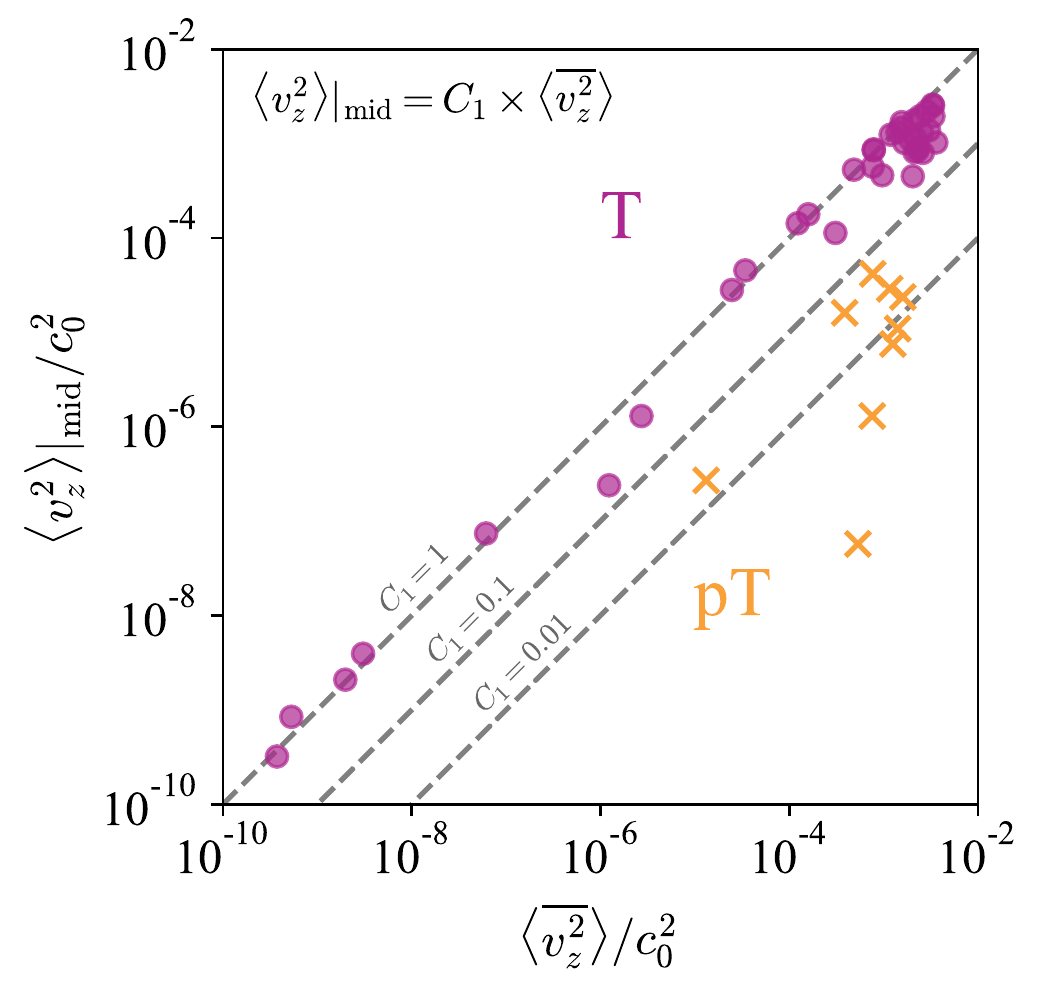}
        \end{center}
        \caption{Time and vertical mean squared vertical velocity, $\langle \overline{v_z^2}\rangle$ vs. time mean squared vertical velocity at the midplane, $\langle v_z^2 \rangle|_{\rm mid}$, at $R=1.0$ for all runs presented in this study. The symbols correspond to the T states (circles) and pT states (crosses). The dashed lines show $\langle v_z^2 \rangle|_{\rm mid} = C_1 \times\langle \overline{v_z^2}\rangle$ with $C_1 = 1$, $0.1$, and $0.01$.}
        \label{fig:vz2_mid_intz_R1}
    \end{figure}
    
    We define the T and pT states more quantitatively by using the time-averaged $v_z^2$ at the midplane, $\langle v_z^2 \rangle|_{\rm mid}$, and $v_z^2$ averaged both in time and in the vertical direction, $\langle \overline{v_z^2}\rangle$.
    The ratio between the two quantities reflects how strongly the vertical gas flow penetrates into the midplane region. 
    Simulations exhibiting strongly overshooting vertical gas flows (T states) yield $\langle v_z^2 \rangle|_{\rm mid} \approx \langle\overline{v_z^2}\rangle$, whereas those with a less turbulent midplane region than in the unstable layers (pT states) yield $ \langle v_z^2 \rangle|_{\rm mid} \ll \langle\overline{v_z^2}\rangle$.
    Figure \ref{fig:vz2_mid_intz_R1} shows $\langle v_z^2 \rangle|_{\rm mid}$ versus $\langle\overline{v_z^2} \rangle$ of all runs presented in this study.
    A majority of our simulations result in either $\langle v_z^2 \rangle|_{\rm mid} \approx \langle \overline{v_z^2}\rangle$ or $\langle v_z^2 \rangle|_{\rm mid} \approx 0.01$--$0.1\langle \overline{v_z^2}\rangle$.
    In the following, we refer to the T and pT states as saturated states with $\langle v_z^2 \rangle|_{\rm mid} > 0.1\langle \overline{v_z^2}\rangle$ and $\langle v_z^2 \rangle|_{\rm mid} < 0.1\langle \overline{v_z^2}\rangle$, respectively.
    The states of all our simulation runs are summarized in table \ref{t:beta} of appendix \ref{appendix:parameter_sets}.
    
    \begin{figure}
        \begin{center}
        \includegraphics[width=80mm,bb = 0 0 370 597]{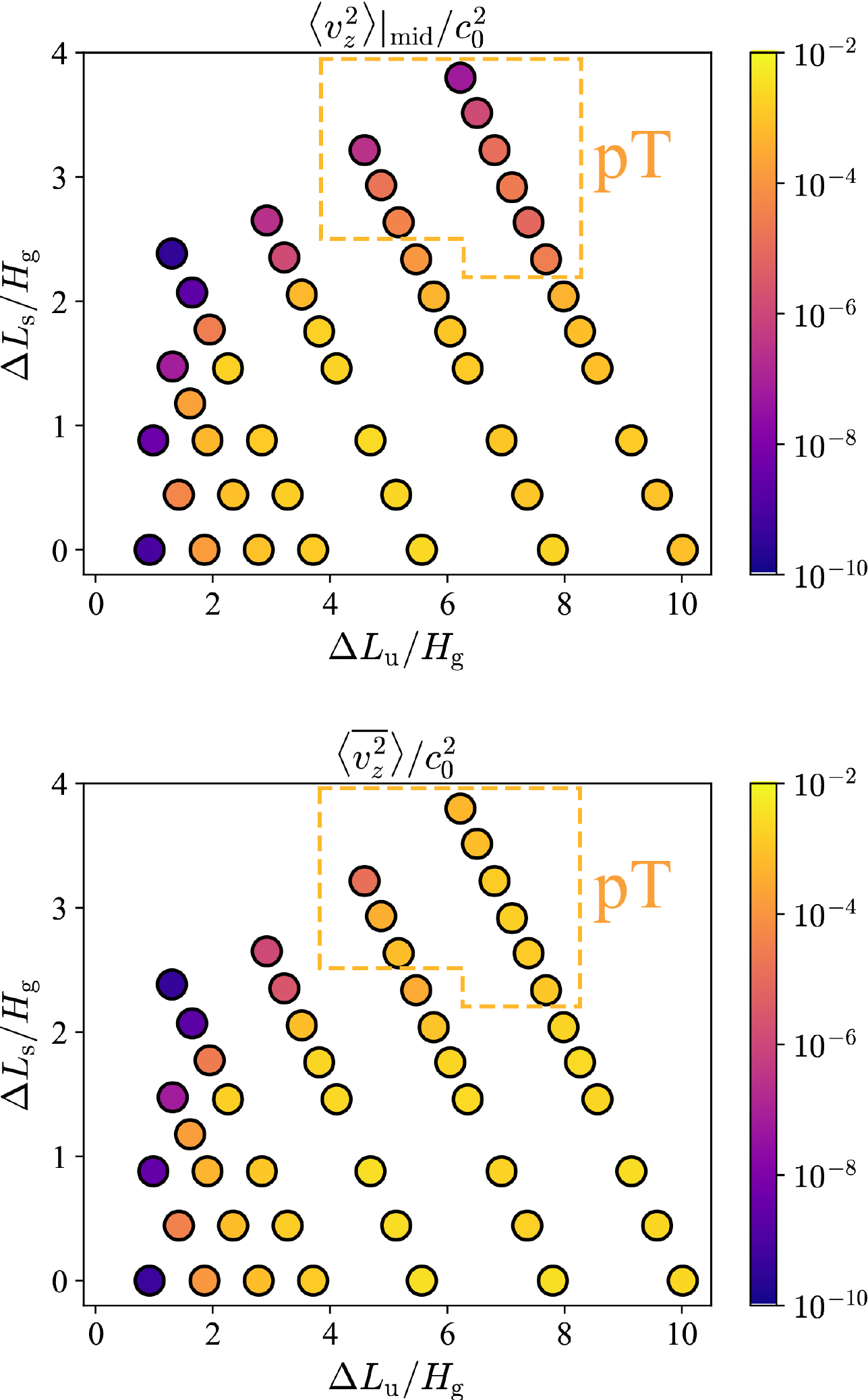}
        \end{center}
        \caption{Time mean squared vertical velocity at the midplane $\langle v_z^2 \rangle|_{\rm mid}$ (upper panel) and its vertical average $\langle \overline{v_z^2}\rangle$ (lower panel) at $R=1.0$ from all simulation runs, mapped in the $\Delta L_{\rm u}$--$\Delta L_{\rm s}$ plane. The dashed line indicates the simulations relaxing to pT states ($\langle v_z^2 \rangle|_{\rm mid} < 0.1\langle \overline{v_z^2}\rangle$). }
        \label{fig:vz2_mid_int}
    \end{figure}    
    
    From the examples shown in figures \ref{fig:vz_colormap_Rz} and \ref{fig:vz_colormap_tz}, one can expect that the thickness of the linearly VSI-stable layer, $\Delta L_{\rm s}$, determines the final saturated state.
    To test this hypothesis, we map in figure \ref{fig:vz2_mid_int} the values of the turbulence diagnostics $\langle v_z^2 \rangle|_{\rm mid}$ and $\langle \overline{v_z^2}\rangle$ from all our simulations against $\Delta L_{\rm u}$ and $\Delta L_{\rm s}$.
    We find that pT states ($\langle v_z^2 \rangle|_{\rm mid} < 0.1\langle \overline{v_z^2}\rangle$) are realized when the linearly stable midplane layer is as wide as $\Delta L_{\rm s} \gtrsim 2H_{\rm g}$.
    
    Another important finding from figure \ref{fig:vz2_mid_int} is that the thickness of the linearly unstable layer, $\Delta L_{\rm u}$, determines the vertically averaged saturation level $\langle \overline{v_z^2}\rangle$.
    Turbulence is largely suppressed at all heights in the cases of $\Delta L_{\rm u} \lesssim 2H_{\rm g}$.
    One can see that $\langle \overline{v_z^2}\rangle$ as well as $\langle v_z^2 \rangle|_{\rm mid}$ decreases sharply from $\sim 2\times 10^{-3}c_0^2$ to $\ll 10^{-3}c_0^2$ as $\Delta L_{\rm u}$ falls below $2H_{\rm g}$.
    
    \begin{figure}
        \begin{center}
        \includegraphics[width=80mm,bb = 0 0 301 284]{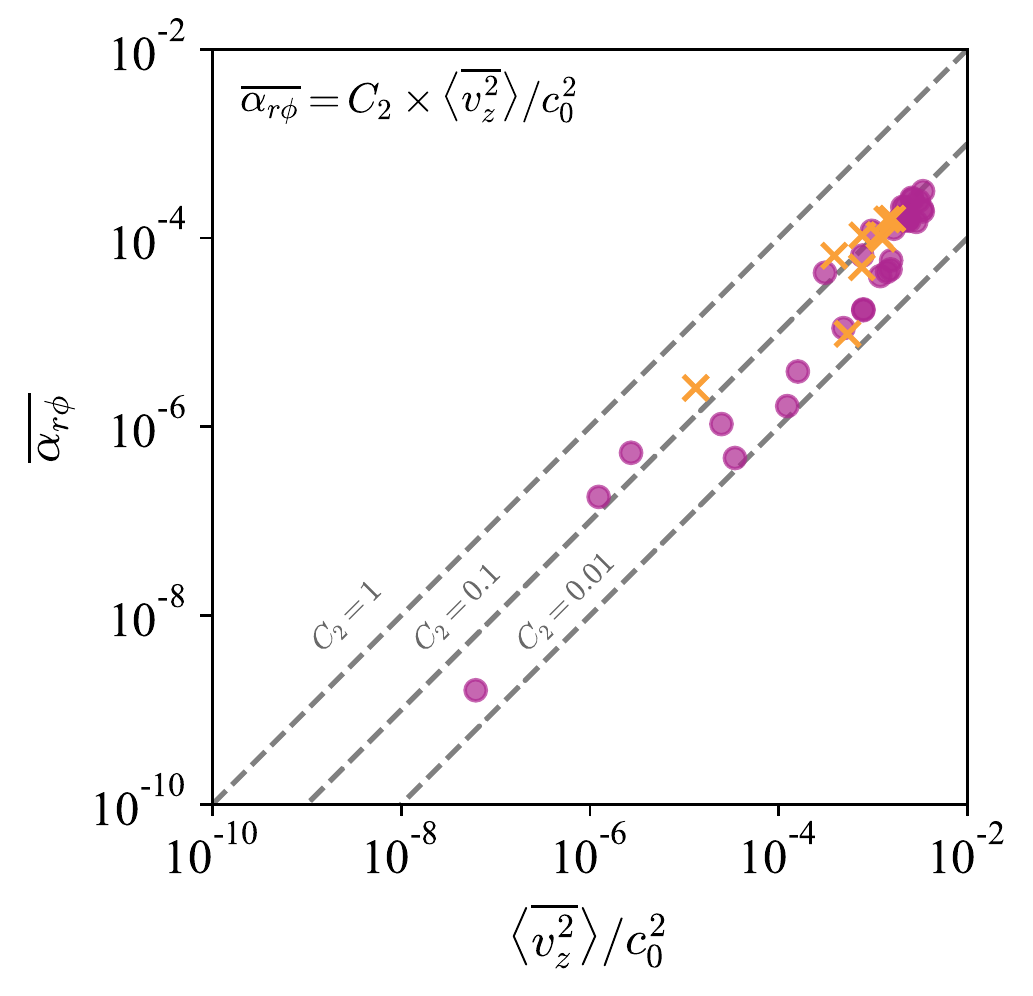}
        \end{center}
        \caption{Time and vertical mean squared vertical velocity, $\langle \overline{v_z^2}\rangle$, vs. vertical mean Reynolds stress, $\overline{\alpha_{r\phi}}$, at $R=1.0$ for all runs. The symbols correspond to the T states (circle) and pT states (crosses). The dashed lines show $\overline{\alpha_{r\phi}} = C_2\times \langle \overline{v_z^2}\rangle/c_0^2$ with $C_2 = 1$, $0.1$, and $0.01$.}
        \label{fig:vz2_intz_alpha}
    \end{figure}
    
    For both the T and pT states, the vertical averaged Reynolds stress $\overline{\alpha_{r\phi}}$ is tightly correlated with $\langle \overline{v_z^2}\rangle$.
    Figure \ref{fig:vz2_intz_alpha} plots $\langle \overline{v_z^2}\rangle$ versus $\overline{\alpha_{r\phi}}$ at $R=1.0$ from all runs presented in this study.
    This figure shows that most simulations produce $\overline{\alpha_{r\phi}} = 0.01$--$0.1\langle \overline{v_z^2}\rangle/c_0^2$.
    
    Our simulation results may change if we perform a three-dimensional (3D) simulation.
    However, the 3D effecs may be minor because \citet{PfeilKlahr:2021nr} already showed that the average vertical velocity in a 3D simulation is comparable to that in 2D simulations.
    
    \subsection{Empirical formulas for turbulent quantities}\label{subsec:fitting}
    
    The results presented in the previous subsection show that important turbulent quantities $\langle v_z^2 \rangle|_{\rm mid}$, $\langle \overline{v_z^2}\rangle$, and $\overline{\alpha_{r\phi}}$ are all tightly correlated with $\Delta L_{\rm s}$ and $\Delta L_{\rm u}$.
    This suggests that one can predict these quantities for general cases as a function of $\Delta L_{\rm s}$ and $\Delta L_{\rm u}$ without having to perform further simulations. 
    Here, we construct such formulas based on our simulation results.
    
    Because there are two types of saturated states, we consider a fitting function for $\langle v_z^2 \rangle|_{\rm mid}$ of the form
    \begin{equation}\label{eq:fitting_vz2}
        \left.\frac{\langle v_z^2 \rangle}{c_{\rm s}^2}\right|_{\rm mid} = f_{\rm T}(\Delta L_{\rm u},~\Delta L_{\rm s}) + f_{\rm pT}(\Delta L_{\rm u},~\Delta L_{\rm s}),
    \end{equation}
    where $f_{\rm T}$ and $f_{\rm pT}$ represent $\langle v_z^2 \rangle|_{\rm mid}$ for the T and pT states, respectively.
    For the T states, figure \ref{fig:vz2_mid_int} shows that $\langle v_z^2 \rangle|_{\rm mid}$ is approximately constant except at $\Delta L_{\rm s} \sim 2H_{\rm g}$, which is the boundary between the T and pT states, and at $\Delta L_{\rm u} \sim 2H_{\rm g}$, where $\langle v_z^2 \rangle|_{\rm mid}$ sharply drops. 
    We reproduce these features with a simple function
    \begin{eqnarray}\label{eq:fitting_vz2_fT}
        f_{\rm T} = 2\times 10^{-3} \varsigma_{3.5}\left(x_1\right) \varsigma_{25}\left(x_2\right),
    \end{eqnarray}
    where $\varsigma_{c}(x)$ is the sigmoid function defined by
    \begin{equation}
        \varsigma_c(x) = \frac{1}{1+\exp{(-cx)}} = \frac{\tanh{(cx/2)}+1}{2}.
    \end{equation}
    The sigmoid function approaches unity and zero in the limits of $x \gg 0$ and $x \ll 0$, respectively.
    Therefore, we adopt this function to reproduce the characteristics of the turbulence diagnostics that are almost constant within the region of $\Delta L_{\rm u} > 2H_{\rm g}$ and $\Delta L_{\rm s} < 2H_{\rm g}$ and change sharply and continuously at some boundaries.
    The arguments of the first and second sigmoid functions in $f_{\rm T}$ express the boundaries that $\langle v_z^2 \rangle|_{\rm mid}$ decreases sharply around $\Delta L_{\rm u} \sim 2H_{\rm g}$ and $\Delta L_{\rm s} \sim 2H_{\rm g}$, respectively.
    We set $x_1 = 3.5\Delta \tilde{L}_{\rm u}-\Delta \tilde{L}_{\rm s}-4.8$ and $x_2 = 0.07\Delta \tilde{L}_{\rm u}-\Delta \tilde{L}_{\rm s}+1.8$, where $\Delta \tilde{L}_{\rm s} = \Delta L_{\rm s}/H_{\rm g}$ and $\Delta \tilde{L}_{\rm u} = \Delta L_{\rm u}/H_{\rm g}$.
    For $f_{\rm pT}$, we use 
    \begin{equation}\label{eq:fitting_vz2_fpT}
        f_{\rm pT} = 2\times 10^{-5} \varsigma_{7}\left(x_3\right),
    \end{equation}
    where $x_3 = \ln(\max\{\Delta \tilde{L}_{\rm u}-2.5,~0\})-\Delta \tilde{L}_{\rm s}+1.8$ that expresses the curve boundary of the drop in the pT states toward large $\Delta L_{\rm s}$.
    We use the natural logarithmic function in $x_3$ to represent this curve boundary.
    
    \begin{figure}
        \begin{center}
        \includegraphics[width=80mm,bb = 0 0 523 843]{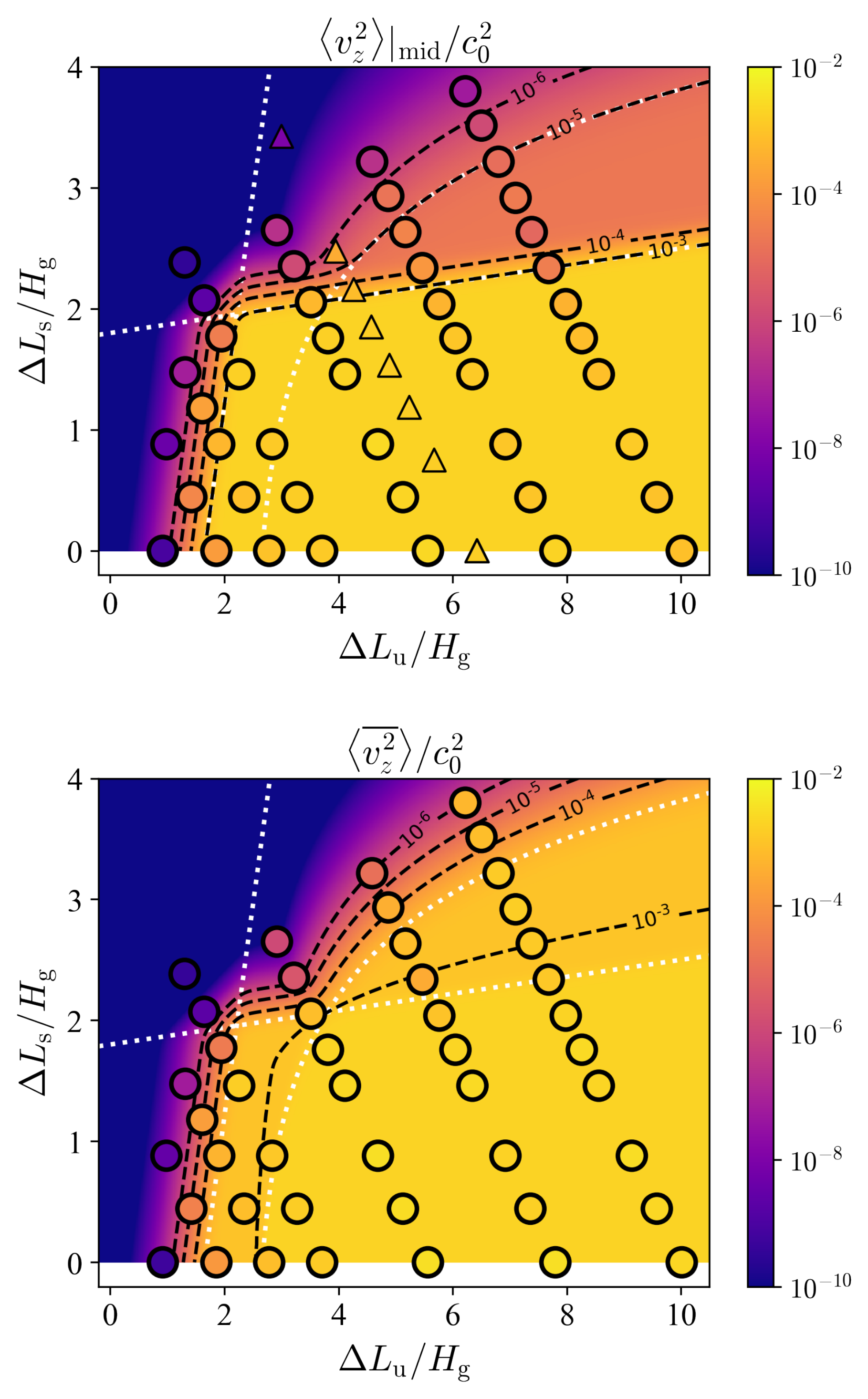}
        \end{center}
        \caption{Upper panel: comparison of $\langle v_z^2 \rangle|_{\rm mid}$ from simulations (points) and the empirical formula (equations \eqref{eq:fitting_vz2}--\eqref{eq:fitting_vz2_fpT}; background) on the $\Delta L_{\rm u}$--$\Delta L_{\rm s}$ plane. The triangles show the simulation results of \citet{PfeilKlahr:2021nr}. The dashed lines are contours of $\langle v_z^2 \rangle|_{\rm mid} = 10^{-3}$, $10^{-4}$, $10^{-5}$, and $10^{-6}$ from the formula. The dotted lines show $3.5\Delta \tilde{L}_{\rm u}-\Delta \tilde{L}_{\rm s}-4.8 = 0$, $0.07\Delta \tilde{L}_{\rm u}-\Delta \tilde{L}_{\rm s}+1.8 = 0$, and $\ln(\max\{\Delta \tilde{L}_{\rm u}-2.5,~0\})-\Delta \tilde{L}_{\rm s}+1.8 = 0$. Lower panel: same as the upper panel, but comparing $\langle \overline{v_z^2}\rangle$ from the simulations and from equations \eqref{eq:fitting_vz2_int}--\eqref{eq:fitting_vz2_int_gpT}. }
        \label{fig:fit_vz2_mid}
    \end{figure}
    
    \begin{figure}
        \begin{center}
        \includegraphics[width=80mm,bb = 0 0 407 280]{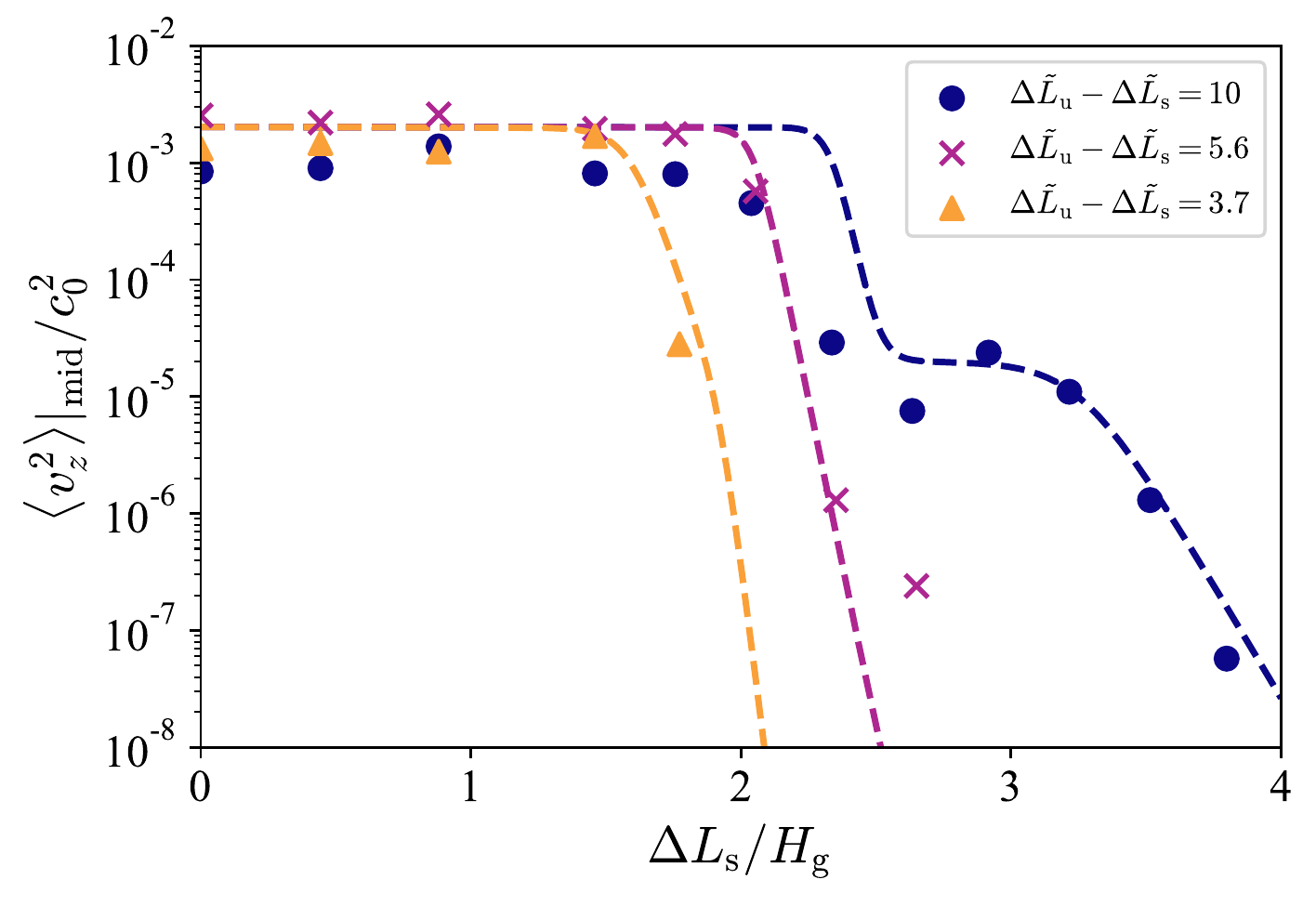}
        \end{center}
        \caption{Comparison of $\langle v_z^2\rangle_{\rm mid}$ from simulations  and empirical formula (equations~\eqref{eq:fitting_vz2}--\eqref{eq:fitting_vz2_fpT}; dashed lines) as a function of $\Delta L_{\rm s}$ The circles, crosses, and triangles are from simulations with $\Delta \tilde{L}_{\rm u}-\Delta \tilde{L}_{\rm s} = 10.0$, $5.6$, and $3.7$, respectively.}
        \label{fig:vz2_mid_simu_fit_Lu_line}
    \end{figure}

    The upper panel of figure \ref{fig:fit_vz2_mid} compares the values of $\langle v_z^2 \rangle|_{\rm mid}$ from our simulations with those from the empirical formula.
    This formula reproduces the simulation results of $\langle v_z^2 \rangle|_{\rm mid}/c_0^2 \sim 10^{-3}$ (T state) and $\sim 10^{-5}$ (pT state).
    The sigmoid functions in $f_{\rm T}$ and $f_{\rm pT}$ are also useful to represent the sharp decrease in $\Delta L_{\rm u} \lesssim 2H_{\rm g}$ and $\Delta L_{\rm s} \sim 2H_{\rm g}$.
    Furthermore, this formula also represents the simulation results of \citet{PfeilKlahr:2021nr}.
    To quantify the errors between the simulation results and formula, we show in figure \ref{fig:vz2_mid_simu_fit_Lu_line} the values of $\langle v_z^2 \rangle|_{\rm mid}$ for some simulations and their corresponding values for the empirical formula as a function of $\Delta L_{\rm s}$.
    This figure indicates that the formula is accurate to less than an order of magnitude in $\langle v_z^2 \rangle|_{\rm mid}$ at $\Delta L_{\rm s} \lesssim 2 H_{\rm g}$.
    Furthermore, the formula replicates the sharp drop at $\Delta L_{\rm s} > 2H_{\rm g}$ and $\langle v_z^2 \rangle|_{\rm mid}/c_0^2 \sim 10^{-5}$ for the pT state's simulations of $\Delta \tilde{L}_{\rm u}-\Delta \tilde{L}_{\rm s} = 10.0$.
    
    To verify the accuracy of the formula, we calculate the root-mean-squared error of the formula with respect to $\log_{10}\langle v_z^2 \rangle|_{\rm mid}$.
    For simulations with $\langle v_z^2 \rangle|_{\rm mid} > 10^{-5}$, the error is 0.48 dex, which means that the formula has an accuracy of less than an order of magnitude.
    The error increases to 0.95 dex if we include all simulations.
    
    Similarly, we propose an empirical formula for $\langle \overline{v_z^2}\rangle$ given by
    \begin{equation}\label{eq:fitting_vz2_int}
        \frac{\langle \overline{v_z^2}\rangle}{c_{\rm s}^2} = g_{\rm T}(\Delta L_{\rm u},~\Delta L_{\rm s}) + g_{\rm pT}(\Delta L_{\rm u},~\Delta L_{\rm s}),
    \end{equation}
    where $g_{\rm T}$ and $g_{\rm pT}$ represent $\langle \overline{v_z^2}\rangle$ in the T and pT states, respectively.
    Because the difference between $\langle v_z^2 \rangle|_{\rm mid}$ and $\langle \overline{v_z^2}\rangle$ appears in only the pT state (see figure \ref{fig:vz2_mid_int}), we fit $\langle \overline{v_z^2}\rangle$ by varying only the coefficients of $f_{\rm T}$ and $f_{\rm pT}$.
    Therefore, we determine $g_{\rm T}$ and $g_{\rm pT}$ as
    \begin{equation}\label{eq:fitting_vz2_int_gT}
        g_{\rm T}(\Delta L_{\rm u},~\Delta L_{\rm s}) = 0.5\cdot f_{\rm T}(\Delta L_{\rm u},~\Delta L_{\rm s}),
    \end{equation}
    \begin{equation}\label{eq:fitting_vz2_int_gpT}
        g_{\rm pT}(\Delta L_{\rm u},~\Delta L_{\rm s}) = 50\cdot f_{\rm pT}(\Delta L_{\rm u},~\Delta L_{\rm s}).
    \end{equation}
    The lower panel of figure \ref{fig:fit_vz2_mid} compares $\langle \overline{v_z^2}\rangle$ from the formula with those from the simulations.
    The formula for $\langle \overline{v_z^2}\rangle$ reproduces the high turbulence level ($\langle \overline{v_z^2}\rangle \sim 10^{-3}c_0^2$) and a sharp decrease at $\Delta L_{\rm u} \lesssim 2H_{\rm g}$.
    
    \begin{figure}
        \begin{center}
        \includegraphics[width=80mm,bb = 0 0 396 274]{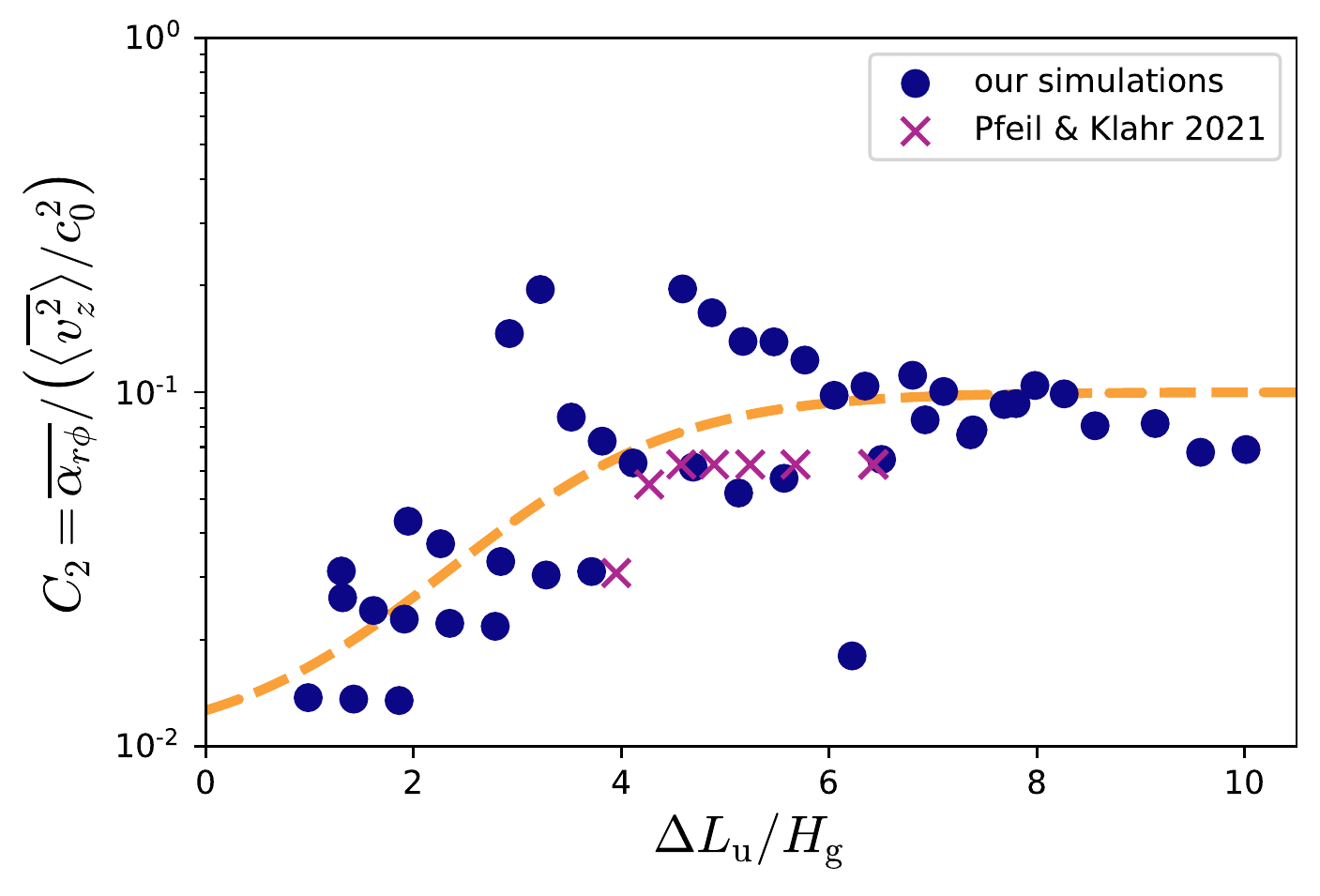}
        \end{center}
        \caption{Ratio of $\overline{\alpha_{r\phi}}$ and $\langle \overline{v_z^2}\rangle/c_0^2$, $C_2$, as a function of $\Delta L_{\rm u}$. The points and crosses plot the simulation results for all runs presented in this study and for \citet{PfeilKlahr:2021nr}, respectively. The dashed line shows the fitting function in equation \eqref{eq:C2_fit}.}
        \label{fig:C2}
    \end{figure}
    
    \begin{figure}
        \begin{center}
        \includegraphics[width=80mm,bb = 0 0 404 280]{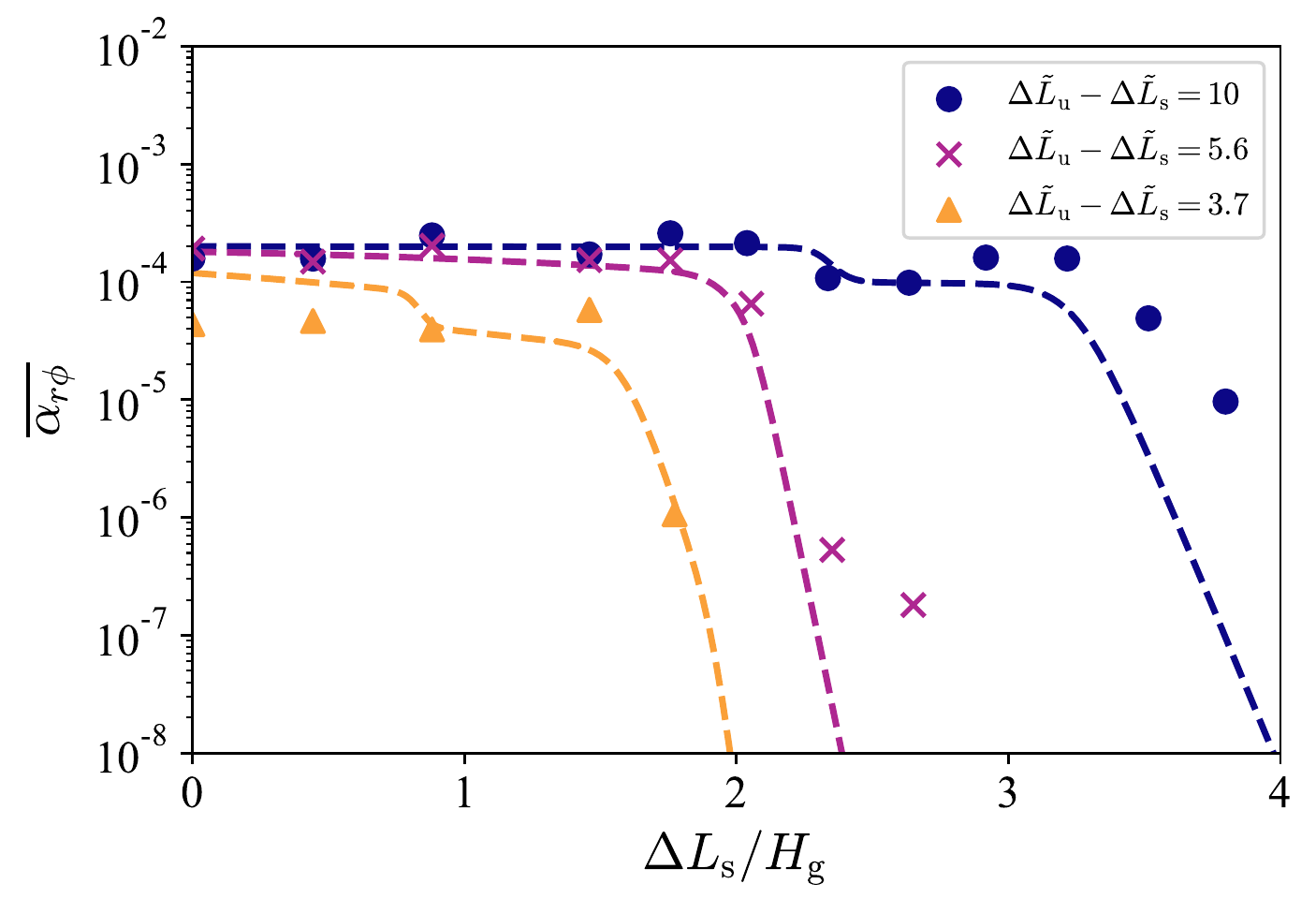}
        \end{center}
        \caption{Comparison of $\overline{\alpha_{r\phi}}$ from simulation results (circles, crosses, and triangles) and empirical formula (equation \eqref{eq:fitting_alpha}; dashed lines) as a function of $\Delta L_{\rm s}$. The circles, crosses, and triangles are from simulations with $\Delta \tilde{L}_{\rm u}-\Delta \tilde{L}_{\rm s} = 10.0$, $5.6$, and $3.7$, respectively.}
        \label{fig:alpha_rphi_simu_fit_Ls_line_fit}
    \end{figure}
    
    The tight correlation between $\overline{\alpha_{r\phi}}$ and $\langle \overline{v_z^2}\rangle$ (figure \ref{fig:vz2_intz_alpha}) motivates us to consider a formula for $\overline{\alpha_{r\phi}}$ of the form
    \begin{eqnarray}\label{eq:fitting_alpha}
        \overline{\alpha_{r\phi}} &=&  C_2\left(\Delta L_{\rm u}\right)\frac{\langle \overline{v_z^2}\rangle}{c_0^2} \nonumber \\
        &=& C_2\left(\Delta L_{\rm u}\right)\left[g_{\rm T}(\Delta L_{\rm u},~\Delta L_{\rm s}) + g_{\rm pT}(\Delta L_{\rm u},~\Delta L_{\rm s}) \right],
    \end{eqnarray}
    where $C_2\left(\Delta L_{\rm u}\right)$ represents the ratio of $\overline{\alpha_{r\phi}}$ to $\langle \overline{v_z^2}\rangle/c_0^2$.
    Based on the simulation results shown in figure~\ref{fig:vz2_intz_alpha}, we propose
    \begin{equation}\label{eq:C2_fit}
        C_2\left(\Delta L_{\rm u}\right) = 0.01 + 0.09\varsigma_1 \left( \Delta \tilde{L}_{\rm u} - 3.5 \right).
    \end{equation}
    This function smoothly decreases from $0.01$ to $0.1$ as $\Delta L_{\rm u}$ decreases.
    Figure \ref{fig:C2} shows that equation~\eqref{eq:C2_fit} reproduces the ratio $\overline{\alpha_{r\phi}}/(\langle \overline{v_z^2}\rangle/c_0^2)$ from the our simulations as well as simulations of \citet{PfeilKlahr:2021nr} to within one order of magnitude.
    The root-mean-squared error of the formula for $\log_{\rm 10}[\overline{\alpha_{r\phi}}/(\langle \overline{v_z^2}\rangle/c_0^2)]$ is 0.24 dex.
    If we limit the comparison to cases with $\overline{\alpha_{r\phi}} \gtrsim 10^{-5}$, the formula is accurate to within a factor of a few. 
    This can be seen in figure \ref{fig:alpha_rphi_simu_fit_Ls_line_fit}, where we plot the values of $\overline{\alpha_{r\phi}}$ from the simulations and the formula as a function of $\Delta L_{\rm s}$.
    
\section{Discussion}\label{sec:discussion}
    \subsection{Estimating dust vertical diffusion coefficient}\label{subsec:estimate_vartical_diffusion}
    The predominantly vertical gas motion in VSI-driven turbulence causes strong vertical dust diffusion \citep{FlockNelson+2017,Flock:2020aa}.
    Our simulations show that the gas velocity dispersion at the midplane varies with the thicknesses of the VSI-stable and unstable layers.   
    Qualifying how the dust diffusion coefficient varies will be useful for studying planetesimal formation (e.g., \citealt{Johansen:2009aa}) and testing theory with millimeter observations of dust rings and gaps (e.g., \citealt{Pinte:2016aa}).
    A direct measurement of the dust vertical diffusion coefficient requires a calculation of dust grains' motion in hydrodynamical simulations, which is beyond the scope of this study.
    Here, we indrectly estimate the dust vertical diffusion coefficient using the gas vertical velocity dispersion measured in our simulations.
    
    Formally, the dust vertical diffusion coefficient is defined as 
    \begin{equation}
        D_z = \frac{1}{2}\frac{d\langle z^2 \rangle_{\rm p}}{dt},
    \end{equation}
    where $\langle z^2 \rangle_{\rm p}$ is the ensemble average of $z^2$ for dust grains.
    For grains whose stopping time is shorter than the orbital period and the correlation time $\tau_{\rm corr}$ of turbulence, $D_z$ reduces to the gas vertical diffusion coefficient, which can be estimated as \citep{FromangPapaloizou:2006rz,YoudinLithwick2007}
    \begin{equation}
        D_z \sim \langle v_z^2 \rangle \tau_{\rm corr}.
    \end{equation}
    It is useful to normalize $D_z$ as 
    \begin{equation}\label{eq:alpha_z}
        \alpha_z \equiv \frac{D_z}{c_{\rm s} H_{\rm g}} \sim   \frac{\langle v_z^2 \rangle}{c_{\rm s}^2}\cdot \tau_{\rm corr}\Omega_{\rm K},
    \end{equation}
    where we have used $H_{\rm g} = c_{\rm s}/\Omega_{\rm K}$.
    \citet{StollKley:2016vp} and \citet{Flock:2020aa} independently estimate $\tau_{\rm corr}$ and report  $\tau_{\rm corr}\Omega_{\rm K} \sim 0.2$ and $\sim 20$, respectively\footnote{\citet{Flock:2020aa} report a dimensionless vertical diffusion coefficient and a mean squared vertical velocity of $\alpha_z = 5.4 \times 10^{-3}$ and $\langle v_z^2 \rangle^{1/2} = 0.0166c_{\rm s}$, respectively. These values and the relation in equation \eqref{eq:alpha_z} lead to $\tau_{\rm corr}\Omega_{\rm K} \sim 20$.}. 
    
    We are particularly interested in the value of $\alpha_z$ at the midplane, $\alpha_{z,\rm mid}$, because the thickness of the dust sedimentary layer at the midplane scales as $\alpha_{z,\rm mid}^{-1/2}$ \citep{Dubrulle+1995,YoudinLithwick2007}.
    Our simulations show $\langle v_z^2 \rangle|_{\rm mid}/c_{\rm s}^2 \sim 2\times 10^{-3}$ (section \ref{subsec:overview}) for fully developed VSI-driven turbulence in T states with $\Delta  L_{\rm u} \gtrsim 2H_{\rm g}$.
    For this case, equation \eqref{eq:alpha_z} predicts a dimensionless vertical diffusion coefficient of $\alpha_{z, \rm mid} \sim 4 \times 10^{-4\dots-2}$, with the uncertainty originating from that of $\tau_{\rm corr}$.
    This predicted value is higher than the dimensionless Reynolds stress in VSI-driven turbulence ($\alpha_{r\phi}\sim 2\times 10^{-4}$), reflecting the predominantly vertical motion of VSI-driven turbulence.
    For the cases of $\Delta L_{\rm u} < 2H_{\rm g}$ and $\Delta L_{\rm s} > 2H_{\rm g}$, equation~\eqref{eq:alpha_z} predicts much smaller diffusion coefficients of $\alpha_{z, \rm mid} \ll 10^{-4}$ and $\alpha_{z, \rm mid} \sim 4 \times 10^{-6\dots-4}$, respectively.
    Implications of the suppressed VSI-driven turbulence for dust evolution and disk observations are discussed in the following section. 
    
    \subsection{Implications for dust evolution and observations of protoplanetary disks}\label{subsec:implication}
    
    The suppression of VSI-driven turbulence at the midplane in the cases of $\Delta L_{\rm u} < 2H_{\rm g}$ or $\Delta L_{\rm s} > 2H_{\rm g}$ has important implications for dust settling and planetesimal formation in the outer regions of protoplanetary disks.
    Weak turbulence yields low relative velocities of dust particles, which is preferred for dust growth through coagulation without collisional fragmentation and erosion (e.g., \citealt{Brauer:2008aa,Okuzumi:2012aa}).
    Furthermore, the weak turbulent diffusion ($\alpha_z \lesssim 10^{-4}$) in suppressed turbulence would promote planetesimal formation through the streaming and gravitational instabilities (e.g., \citealt{Sekiya:1998aa,Youdin:2002aa,Johansen:2009aa,GoleSimon+:2020aa,UmurhanEstrada+:2020yi,ChenLin:2020kh}).
    Therefore, disk regions with $\Delta L_{\rm u} < 2H_{\rm g}$ or $\Delta L_{\rm s} > 2H_{\rm g}$ would be preferential sites for planetesimal formation.
    
    The suppression of VSI-driven turbulence may also explain the low level of vertical dust diffusion inferred from ALMA observations of some protoplanetary disks.
    The dust rings around HL Tau and Oph 163131 exhibit well-separated morphology in millimeter images, indicating that the large dust particles in the rings have settled onto the midplane.
    Assuming that millimeter-sized particles dominate the millimeter emission, these observations point to small vertical diffusion coefficients of $\alpha_z \sim$ a few $10^{-4}$ for HL Tau \citep{Pinte:2016aa} and of $\alpha \lesssim 10^{-5}$ for Oph 163131 \citep{VillenaveStapelfeldt+:2022pp}.
    These estimated values of $\alpha_z$ are consistent with suppressed VSI-driven turbulence in the cases of $\Delta L_{\rm u} < 2H_{\rm g}$ or $\Delta L_{\rm s} > 2H_{\rm g}$ (see section \ref{subsec:estimate_vartical_diffusion}).
    Therefore, we hypothesize that VSI-driven turbulence is indeed suppressed in the outer regions of these disks.
    Testing this hypothesis requires detailed modeling of these disks' cooling structure.
    
    \subsection{Need for self-consistent modeling of dust and VSI evolution}
    Because dust particles control disk cooling, whether the condition $\Delta L_{\rm u} < 2H_{\rm g}$ or $\Delta L_{\rm s} > 2H_{\rm g}$ for the suppression of VSI-driven turbulence is realized would depend on the size and amount of the particles.
    Dust growth and settling lead to a VSI-unstable region that has smaller $\Delta L_{\rm u}$, i.e., that is more confined around the midplane \citep{FukuharaOkuzumi+:2021ca}.
    Depletion of small grains that dominate the gas disk cooling can also increase the cooling time \citep{DullemondZiampras+:2022aa} and thereby make the VSI-unstable region smaller.
    These effects may result in suppression of VSI-driven turbulence. 
    On the other hand, an increase in the dust surface density leads to a wider optically thick region around the midplane. 
    This may make the region around the midplane cooling less efficient and consequently suppress VSI-driven turbulence at the midplane. 
    Assessing whether $\Delta L_{\rm u} < 2H_{\rm g}$ or $\Delta L_{\rm s} > 2H_{\rm g}$ can be realized under realistic conditions requires self-consistent modeling of dust growth, dust surface density evolution,  and disk cooling that takes into account gas--dust thermal coupling.

    Moreover, because a change in the saturated state of VSI-driven turbulence would also affect dust evolution, the evolution of dust and VSI should be coupled.
    For instance, suppression of VSI-driven turbulence by dust growth, if it really occurs, would reduce the collision velocity between the dust particles and thus promote their further growth. 
    A similar positive feedback can also be expected for suppression of VSI-driven turbulence by dust settling.  
    These positive feedback effects can be important for understanding planet formation and turbulence in outer disk regions.
    
    To quantify these effects, the empirical formulas presented in section \ref{subsec:fitting} will be useful.
    These formulas represent the correlation between the thicknesses of the VSI-unstable and stable layers and the VSI-driven turbulence intensity.
    The size of dust particles as well as the spatial distribution controls gas cooling and determines the thicknesses of the VSI-unstable and stable layers.
    Our formulas can be used to predict how the saturated level of VSI-driven turbulence would evolve with the long-term evolution of dust.
    This will be done in our future work.
    
    \subsection{Limitation of our simulations} \label{subsec:limitation}
    Our simulations are subject to two important limitations that should be addressed in future work. 
    First, the cooling rates adopted in our simulations are vertically varying but constant in time.
    In reality, because VSI-driven turbulence alters the dust distribution (e.g., \citealt{Flock:2020aa}), the cooling time distribution determined by dust can change with turbulence.
    Therefore, turbulence and cooling time would evolve simultaneously until turbulence and dust profile are in equilibrium states.
    If the dust evolution timescale is longer than the timescale for VSI-driven turbulence saturation, we can investigate the co-evolution of dust and turbulence using empirical formulas presented in section \ref{subsec:fitting}.
    This is because it is reasonable to assume that the size and spatial distribution of dust grains will not change while the VSI develops turbulence.
    On the other hand, if this assumption breaks down, the stability of this system should be studied in hydrodynamical simulations that include the thermal coupling between gas and dust.
    We plan to address these open issues in future work.
    
    Second, our simulations neglect the effects of dust and magnetic fields on gas disk dynamics.
    Dust would increase the effective buoyancy frequency of the gas \citep{Lin:2017aa} and weaken the VSI \citep{Lin:2019aa}.
    Magnetic fields threading the disk may also suppress the VSI either directly through magnetic tension or indirectly through MRI turbulence \citep{NelsonGresselUmurhan2013,LatterPapaloizou2018,Cui:2020aa}.
    The roles of magnetic fields in VSI suppression can be positive or negative depending on non-ideal magnetohydrodynamical effects (ambipolar diffusion, Ohmic resistivity, and Hall effect;  \citealt{Cui:2020aa,CuiBai:2022aa,CuiLin:2021cj,LatterKunz:2022ic}).
    We plan to quantify these effects using simulations including dust feedback and magnetic field in the future.

\section{Summary}\label{sec:summary}
In this study, we have investigated how the saturated state of VSI-driven turbulence depends on the vertical profile of the disk cooling rate.
We have performed global two-dimensional hydrodynamical simulations of an axisymmetric protoplanetary disk with vertically varying cooling times.
Our key findings are summarized as follows.
\begin{enumerate}
    \item The thickness of the linearly VSI-stable layer at the midplane determines the vertical structure of VSI-driven turbulence in a steady state (figures \ref{fig:vz_colormap_Rz} and \ref{fig:vz_colormap_tz}). We have identified two final saturated states of turbulence. In the first state, the vertical gas motion generated in the linearly VSI-unstable layers penetrates the VSI-stable midplane layer ({\it T states}). In the second state, the vertical gas motion is well confined in the unstable layers ({\it pT states}), leaving the stable midplane layer only weakly turbulent. Using the time averaged squared vertical velocity at the midplane $\langle v_z^2 \rangle|_{\rm mid}$ and $v_z^2$ averaged both in time and in the vertical direction $\langle \overline{v_z^2}\rangle$, we refer to the T and pT states with $\langle v_z^2 \rangle|_{\rm mid} > 0.1\langle \overline{v_z^2}\rangle$ and $\langle v_z^2 \rangle|_{\rm mid} < 0.1\langle \overline{v_z^2}\rangle$, respectively (figure \ref{fig:vz2_mid_intz_R1}).
    \item The pT states are realized when the thickness of the VSI-stable midplane layer $\Delta L_{\rm s}$ is larger than two gas scale heights (figure \ref{fig:vz2_mid_int}). When the thickness of the VSI-unstable layer $\Delta L_{\rm u}$ is thinner than $2H_{\rm g}$, VSI-driven turbulence is also largely suppressed at all heights. The turbulence diagnostic value $\langle \overline{v_z^2}\rangle$ as well as $\langle v_z^2 \rangle|_{\rm mid}$ decreases sharply from $\sim 2\times 10^{-3}c_0^2$ to $\ll 10^{-3}c_0^2$, where $c_0$ is the sound speed, as $\Delta L_{\rm u}$ falls below $2H_{\rm g}$ (figure \ref{fig:vz2_mid_int}).
    \item For both the T and pT states, the vertical averaged Reynolds stress $\overline{\alpha_{r\phi}}$ and $\langle \overline{v_z^2}\rangle$ are connected by $\overline{\alpha_{r\phi}} = 0.01$--$0.1\langle \overline{v_z^2}\rangle/c_0^2$ (figure \ref{fig:vz2_intz_alpha}).
    \item We propose empirical formulas for the turbulence diagnostics $\langle v_z^2 \rangle|_{\rm mid}$, $\langle \overline{v_z^2}\rangle$, and $\overline{\alpha_{r\phi}}$ as a function of $\Delta L_{\rm u}$ and $\Delta L_{\rm s}$ (equations \eqref{eq:fitting_vz2}, \eqref{eq:fitting_vz2_int}, and \eqref{eq:fitting_alpha}, respectively). These formulas reproduce the tightly correlation of turbulence diagnostics with $\Delta L_{\rm u}$ and $\Delta L_{\rm s}$ (figure \ref{fig:fit_vz2_mid}) and have an accuracy of less than an order of magnitude in strong turbulence (figures \ref{fig:vz2_mid_simu_fit_Lu_line} and \ref{fig:alpha_rphi_simu_fit_Ls_line_fit}). These formulas will be useful for predicting how the states of VSI-driven turbulence vary with the long-term evolution of dust.
\end{enumerate}

Our results suggest that the suppression of VSI-driven turbulence at the midplane in the cases of $\Delta L_{\rm u} < 2H_{\rm g}$ or $\Delta L_{\rm s} > 2H_{\rm g}$ can lead to the strongly vertical settling of dust particles.
This effect may promote planetesimal formation through dust coagulation and the gravitational and streaming instabilities in outer disk regions.
This effect may also explain the low level of vertical dust diffusion observed in the dust rings of some protoplanetary disks.

For simplicity, the present study has modeled the vertical cooling rate profile with a parameterized analytic function. 
In reality, gas cooling is regulated by dust particles, and therefore the disks' cooling structure should depend on the particles' size and spatial distribution.
This implies that the evolution of dust and VSI can be coupled because VSI-driven turbulence can affect dust evolution.
For instance, dust growth would make the VSI-unstable region confined around the midplane and consequently suppress VSI-driven turbulence.
This turbulence suppression by dust growth may reduce the collision velocity between dust grains and thus promote their further growth.
To quantify this positive feedback effect, the empirical formulas presented in this study may be useful.
Verification of planetesimal formation by this positive feedback requires studying how the VSI and dust co-evolve in the future.

\begin{ack}
    We thank Mario Flock for a discussion about the correlation time of VSI-driven turbulence.
    We also thank the anonymous referee for helpful comments.
    This work was supported by JSPS KAKENHI Grant Numbers JP18H05438, JP20H00182, JP20H01948, JP20J01376, and JP22J22593.
    Numerical computations were carried out on Cray XC50 at Center for Computational Astrophysics, National Astronomical Observatory of Japan.
\end{ack}

\bibliographystyle{apj}
\bibliography{Fukuhara+22_PASJ}

\begin{thebibliography}{}
\expandafter\ifx\csname natexlab\endcsname\relax\def\natexlab#1{#1}\fi

\bibitem[{{ALMA Partnership} {et~al.}(2015){ALMA Partnership}, {Brogan},
  {P{\'e}rez}, {Hunter}, {Dent}, {Hales}, {Hills}, {Corder}, {Fomalont},
  {Vlahakis}, {Asaki}, {Barkats}, {Hirota}, {Hodge}, {Impellizzeri}, {Kneissl},
  {Liuzzo}, {Lucas}, {Marcelino}, {Matsushita}, {Nakanishi}, {Phillips},
  {Richards}, {Toledo}, {Aladro}, {Broguiere}, {Cortes}, {Cortes}, {Espada},
  {Galarza}, {Garcia-Appadoo}, {Guzman-Ramirez}, {Humphreys}, {Jung}, {Kameno},
  {Laing}, {Leon}, {Marconi}, {Mignano}, {Nikolic}, {Nyman}, {Radiszcz},
  {Remijan}, {Rod{\'o}n}, {Sawada}, {Takahashi}, {Tilanus}, {Vila Vilaro},
  {Watson}, {Wiklind}, {Akiyama}, {Chapillon}, {de Gregorio-Monsalvo}, {Di
  Francesco}, {Gueth}, {Kawamura}, {Lee}, {Nguyen Luong}, {Mangum}, {Pietu},
  {Sanhueza}, {Saigo}, {Takakuwa}, {Ubach}, {van Kempen}, {Wootten},
  {Castro-Carrizo}, {Francke}, {Gallardo}, {Garcia}, {Gonzalez}, {Hill},
  {Kaminski}, {Kurono}, {Liu}, {Lopez}, {Morales}, {Plarre}, {Schieven},
  {Testi}, {Videla}, {Villard}, {Andreani}, {Hibbard}, \&
  {Tatematsu}}]{ALMA+2014}
{ALMA Partnership}, {Brogan}, C.~L., {P{\'e}rez}, L.~M., {et~al.} 2015, \apjl,
  808, L3

\bibitem[{{Andrews} {et~al.}(2018){Andrews}, {Huang}, {P{\'e}rez}, {Isella},
  {Dullemond}, {Kurtovic}, {Guzm{\'a}n}, {Carpenter}, {Wilner}, {Zhang}, {Zhu},
  {Birnstiel}, {Bai}, {Benisty}, {Hughes}, {{\"O}berg}, \&
  {Ricci}}]{Andrews+2018}
{Andrews}, S.~M., {Huang}, J., {P{\'e}rez}, L.~M., {et~al.} 2018, \apjl, 869,
  L41

\bibitem[{{Arlt} \& {Urpin}(2004)}]{ArltUrpin2004}
{Arlt}, R., \& {Urpin}, V. 2004, \aap, 426, 755

\bibitem[{{Bai}(2015)}]{Bai2015}
{Bai}, X.-N. 2015, \apj, 798, 84

\bibitem[{{Balbus} \& {Hawley}(1991)}]{BalbusHawley1991}
{Balbus}, S.~A., \& {Hawley}, J.~F. 1991, \apj, 376, 214

\bibitem[{{Barge} \& {Sommeria}(1995)}]{BargeSommeria:1995qd}
{Barge}, P., \& {Sommeria}, J. 1995, \aap, 295, L1

\bibitem[{{Barranco} {et~al.}(2018){Barranco}, {Pei}, \&
  {Marcus}}]{BarrancoPei+:2018kc}
{Barranco}, J.~A., {Pei}, S., \& {Marcus}, P.~S. 2018, \apj, 869, 127

\bibitem[{{B{\'e}thune} {et~al.}(2017){B{\'e}thune}, {Lesur}, \&
  {Ferreira}}]{BethuneLesur+:2017aa}
{B{\'e}thune}, W., {Lesur}, G., \& {Ferreira}, J. 2017, \aap, 600, A75

\bibitem[{{Brauer} {et~al.}(2008){Brauer}, {Dullemond}, \&
  {Henning}}]{Brauer:2008aa}
{Brauer}, F., {Dullemond}, C.~P., \& {Henning}, T. 2008, \aap, 480, 859

\bibitem[{{Carrera} {et~al.}(2015){Carrera}, {Johansen}, \&
  {Davies}}]{Carrera:2015aa}
{Carrera}, D., {Johansen}, A., \& {Davies}, M.~B. 2015, \aap, 579, A43

\bibitem[{{Chen} \& {Lin}(2020)}]{ChenLin:2020kh}
{Chen}, K., \& {Lin}, M.-K. 2020, \apj, 891, 132

\bibitem[{{Cui} \& {Bai}(2020)}]{Cui:2020aa}
{Cui}, C., \& {Bai}, X.-N. 2020, \apj, 891, 30

\bibitem[{{Cui} \& {Bai}(2021)}]{CuiBai:2021aa}
---. 2021, \mnras, 507, 1106

\bibitem[{{Cui} \& {Bai}(2022)}]{CuiBai:2022aa}
---. 2022, \mnras, 516, 4660

\bibitem[{{Cui} \& {Latter}(2022)}]{CuiLatter:2022aa}
{Cui}, C., \& {Latter}, H.~N. 2022, \mnras, 512, 1639

\bibitem[{{Cui} \& {Lin}(2021)}]{CuiLin:2021cj}
{Cui}, C., \& {Lin}, M.-K. 2021, \mnras, 505, 2983

\bibitem[{{Doi} \& {Kataoka}(2021)}]{DoiKataoka:2021oz}
{Doi}, K., \& {Kataoka}, A. 2021, \apj, 912, 164

\bibitem[{{Drazkowska} {et~al.}(2022){Drazkowska}, {Bitsch}, {Lambrechts},
  {Mulders}, {Harsono}, {Vazan}, {Liu}, {Ormel}, {Kretke}, \&
  {Morbidelli}}]{DrazkowskaBitsch+:2022qi}
{Drazkowska}, J., {Bitsch}, B., {Lambrechts}, M., {et~al.} 2022, arXiv
  e-prints, arXiv:2203.09759

\bibitem[{{Dubrulle} {et~al.}(1995){Dubrulle}, {Morfill}, \&
  {Sterzik}}]{Dubrulle+1995}
{Dubrulle}, B., {Morfill}, G., \& {Sterzik}, M. 1995, \icarus, 114, 237

\bibitem[{{Dullemond} \& {Dominik}(2005)}]{DullemondDominik:2005vy}
{Dullemond}, C.~P., \& {Dominik}, C. 2005, \aap, 434, 971

\bibitem[{{Dullemond} {et~al.}(2022){Dullemond}, {Ziampras}, {Ostertag}, \&
  {Dominik}}]{DullemondZiampras+:2022aa}
{Dullemond}, C.~P., {Ziampras}, A., {Ostertag}, D., \& {Dominik}, C. 2022,
  arXiv e-prints, arXiv:2210.13413

\bibitem[{{Flaherty} {et~al.}(2020){Flaherty}, {Hughes}, {Simon}, {Qi}, {Bai},
  {Bulatek}, {Andrews}, {Wilner}, \& {K{\'o}sp{\'a}l}}]{Flaherty:2020aa}
{Flaherty}, K., {Hughes}, A.~M., {Simon}, J.~B., {et~al.} 2020, \apj, 895, 109

\bibitem[{{Flaherty} {et~al.}(2015){Flaherty}, {Hughes}, {Rosenfeld},
  {Andrews}, {Chiang}, {Simon}, {Kerzner}, \& {Wilner}}]{Flaherty:2015aa}
{Flaherty}, K.~M., {Hughes}, A.~M., {Rosenfeld}, K.~A., {et~al.} 2015, \apj,
  813, 99

\bibitem[{{Flaherty} {et~al.}(2018){Flaherty}, {Hughes}, {Teague}, {Simon},
  {Andrews}, \& {Wilner}}]{Flaherty:2018aa}
{Flaherty}, K.~M., {Hughes}, A.~M., {Teague}, R., {et~al.} 2018, \apj, 856, 117

\bibitem[{{Flaherty} {et~al.}(2017){Flaherty}, {Hughes}, {Rose}, {Simon}, {Qi},
  {Andrews}, {K{\'o}sp{\'a}l}, {Wilner}, {Chiang}, {Armitage}, \&
  {Bai}}]{Flaherty:2017aa}
{Flaherty}, K.~M., {Hughes}, A.~M., {Rose}, S.~C., {et~al.} 2017, \apj, 843,
  150

\bibitem[{{Flock} {et~al.}(2017){Flock}, {Nelson}, {Turner}, {Bertrang},
  {Carrasco-Gonz{\'a}lez}, {Henning}, {Lyra}, \& {Teague}}]{FlockNelson+2017}
{Flock}, M., {Nelson}, R.~P., {Turner}, N.~J., {et~al.} 2017, \apj, 850, 131

\bibitem[{{Flock} {et~al.}(2020){Flock}, {Turner}, {Nelson}, {Lyra}, {Manger},
  \& {Klahr}}]{Flock:2020aa}
{Flock}, M., {Turner}, N.~J., {Nelson}, R.~P., {et~al.} 2020, \apj, 897, 155

\bibitem[{{Flores-Rivera} {et~al.}(2020){Flores-Rivera}, {Flock}, \&
  {Nakatani}}]{Flores-Rivera:2020ab}
{Flores-Rivera}, L., {Flock}, M., \& {Nakatani}, R. 2020, \aap, 644, A50

\bibitem[{{Fricke}(1968)}]{Fricke:1968aa}
{Fricke}, K. 1968, \zap, 68, 317

\bibitem[{{Fromang} \& {Papaloizou}(2006)}]{FromangPapaloizou:2006rz}
{Fromang}, S., \& {Papaloizou}, J. 2006, \aap, 452, 751

\bibitem[{{Fukuhara} {et~al.}(2021){Fukuhara}, {Okuzumi}, \&
  {Ono}}]{FukuharaOkuzumi+:2021ca}
{Fukuhara}, Y., {Okuzumi}, S., \& {Ono}, T. 2021, \apj, 914, 132

\bibitem[{{Goldreich} \& {Schubert}(1967)}]{GS67}
{Goldreich}, P., \& {Schubert}, G. 1967, \apj, 150, 571

\bibitem[{{Goldreich} \& {Ward}(1973)}]{Goldreich:1973aa}
{Goldreich}, P., \& {Ward}, W.~R. 1973, \apj, 183, 1051

\bibitem[{{Gole} {et~al.}(2020){Gole}, {Simon}, {Li}, {Youdin}, \&
  {Armitage}}]{GoleSimon+:2020aa}
{Gole}, D.~A., {Simon}, J.~B., {Li}, R., {Youdin}, A.~N., \& {Armitage}, P.~J.
  2020, \apj, 904, 132

\bibitem[{{Guilloteau} {et~al.}(2012){Guilloteau}, {Dutrey}, {Wakelam},
  {Hersant}, {Semenov}, {Chapillon}, {Henning}, \&
  {Pi{\'e}tu}}]{GuilloteauDutrey+:2012dd}
{Guilloteau}, S., {Dutrey}, A., {Wakelam}, V., {et~al.} 2012, \aap, 548, A70

\bibitem[{{Hughes} {et~al.}(2011){Hughes}, {Wilner}, {Andrews}, {Qi}, \&
  {Hogerheijde}}]{HughesWilner+:2011ed}
{Hughes}, A.~M., {Wilner}, D.~J., {Andrews}, S.~M., {Qi}, C., \& {Hogerheijde},
  M.~R. 2011, \apj, 727, 85

\bibitem[{{Johansen} {et~al.}(2014){Johansen}, {Blum}, {Tanaka}, {Ormel},
  {Bizzarro}, \& {Rickman}}]{Johansen+2014}
{Johansen}, A., {Blum}, J., {Tanaka}, H., {et~al.} 2014, in Protostars and
  Planets VI, ed. H.~{Beuther}, R.~S. {Klessen}, C.~P. {Dullemond}, \&
  T.~{Henning}, 547

\bibitem[{{Johansen} \& {Youdin}(2007)}]{JohansenYoudin2007}
{Johansen}, A., \& {Youdin}, A. 2007, \apj, 662, 627

\bibitem[{{Johansen} {et~al.}(2009){Johansen}, {Youdin}, \& {Mac
  Low}}]{Johansen:2009aa}
{Johansen}, A., {Youdin}, A., \& {Mac Low}, M.-M. 2009, \apjl, 704, L75

\bibitem[{{Kataoka} {et~al.}(2013){Kataoka}, {Tanaka}, {Okuzumi}, \&
  {Wada}}]{Kataoka:2013aa}
{Kataoka}, A., {Tanaka}, H., {Okuzumi}, S., \& {Wada}, K. 2013, \aap, 557, L4

\bibitem[{{Kretke} \& {Lin}(2007)}]{KretkeLin:2007mn}
{Kretke}, K.~A., \& {Lin}, D.~N.~C. 2007, \apjl, 664, L55

\bibitem[{{Latter} \& {Kunz}(2022)}]{LatterKunz:2022ic}
{Latter}, H.~N., \& {Kunz}, M.~W. 2022, \mnras, 511, 1182

\bibitem[{{Latter} \& {Papaloizou}(2018)}]{LatterPapaloizou2018}
{Latter}, H.~N., \& {Papaloizou}, J. 2018, \mnras, 474, 3110

\bibitem[{{Lehmann} \& {Lin}(2022)}]{LehmannLin:2022nr}
{Lehmann}, M., \& {Lin}, M.~K. 2022, \aap, 658, A156

\bibitem[{{Lesur} {et~al.}(2022){Lesur}, {Ercolano}, {Flock}, {Lin}, {Yang},
  {Barranco}, {Benitez-Llambay}, {Goodman}, {Johansen}, {Klahr}, {Laibe},
  {Lyra}, {Marcus}, {Nelson}, {Squire}, {Simon}, {Turner}, {Umurhan}, \&
  {Youdin}}]{LesurErcolano+:2022kp}
{Lesur}, G., {Ercolano}, B., {Flock}, M., {et~al.} 2022, arXiv e-prints,
  arXiv:2203.09821

\bibitem[{{Lin}(2019)}]{Lin:2019aa}
{Lin}, M.-K. 2019, \mnras, 485, 5221

\bibitem[{{Lin} \& {Youdin}(2015)}]{LinYoudin2015}
{Lin}, M.-K., \& {Youdin}, A.~N. 2015, \apj, 811, 17

\bibitem[{{Lin} \& {Youdin}(2017)}]{Lin:2017aa}
---. 2017, \apj, 849, 129

\bibitem[{{Long} {et~al.}(2018){Long}, {Pinilla}, {Herczeg}, {Harsono},
  {Dipierro}, {Pascucci}, {Hendler}, {Tazzari}, {Ragusa}, {Salyk}, {Edwards},
  {Lodato}, {van de Plas}, {Johnstone}, {Liu}, {Boehler}, {Cabrit}, {Manara},
  {Menard}, {Mulders}, {Nisini}, {Fischer}, {Rigliaco}, {Banzatti}, {Avenhaus},
  \& {Gully-Santiago}}]{Long:2018aa}
{Long}, F., {Pinilla}, P., {Herczeg}, G.~J., {et~al.} 2018, \apj, 869, 17

\bibitem[{{Lyra} \& {Umurhan}(2019)}]{LyraUmurhan2019}
{Lyra}, W., \& {Umurhan}, O.~M. 2019, \pasp, 131, 072001

\bibitem[{{Malygin} {et~al.}(2017){Malygin}, {Klahr}, {Semenov}, {Henning}, \&
  {Dullemond}}]{Malygin+2017}
{Malygin}, M.~G., {Klahr}, H., {Semenov}, D., {Henning}, T., \& {Dullemond},
  C.~P. 2017, \aap, 605, A30

\bibitem[{{Manger} {et~al.}(2021){Manger}, {Pfeil}, \&
  {Klahr}}]{MangerPfeil+:2021cm}
{Manger}, N., {Pfeil}, T., \& {Klahr}, H. 2021, \mnras, 508, 5402

\bibitem[{{Mignone} \& {Bodo}(2005)}]{MignoneBodo:2005bv}
{Mignone}, A., \& {Bodo}, G. 2005, \mnras, 364, 126

\bibitem[{{Nakagawa} {et~al.}(1981){Nakagawa}, {Nakazawa}, \&
  {Hayashi}}]{NakagawaNakazawa+:1981wj}
{Nakagawa}, Y., {Nakazawa}, K., \& {Hayashi}, C. 1981, \icarus, 45, 517

\bibitem[{{Nelson} {et~al.}(2013){Nelson}, {Gressel}, \&
  {Umurhan}}]{NelsonGresselUmurhan2013}
{Nelson}, R.~P., {Gressel}, O., \& {Umurhan}, O.~M. 2013, \mnras, 435, 2610

\bibitem[{{Okuzumi} \& {Hirose}(2012)}]{Okuzumi:2012aa}
{Okuzumi}, S., \& {Hirose}, S. 2012, \apjl, 753, L8

\bibitem[{{Okuzumi} {et~al.}(2012){Okuzumi}, {Tanaka}, {Kobayashi}, \&
  {Wada}}]{Okuzumi+2012}
{Okuzumi}, S., {Tanaka}, H., {Kobayashi}, H., \& {Wada}, K. 2012, \apj, 752,
  106

\bibitem[{{Ormel} \& {Cuzzi}(2007)}]{OrmelCuzzi2007}
{Ormel}, C.~W., \& {Cuzzi}, J.~N. 2007, \aap, 466, 413

\bibitem[{{Pfeil} \& {Klahr}(2019)}]{PfeilKlahr2019}
{Pfeil}, T., \& {Klahr}, H. 2019, \apj, 871, 150

\bibitem[{{Pfeil} \& {Klahr}(2021)}]{PfeilKlahr:2021nr}
---. 2021, \apj, 915, 130

\bibitem[{{Pierens}(2021)}]{Pierens:2021aa}
{Pierens}, A. 2021, \mnras, 504, 4522

\bibitem[{{Pinilla} {et~al.}(2012){Pinilla}, {Birnstiel}, {Ricci}, {Dullemond},
  {Uribe}, {Testi}, \& {Natta}}]{PinillaBirnstiel+:2012vz}
{Pinilla}, P., {Birnstiel}, T., {Ricci}, L., {et~al.} 2012, \aap, 538, A114

\bibitem[{{Pinte} {et~al.}(2016){Pinte}, {Dent}, {M{\'e}nard}, {Hales}, {Hill},
  {Cortes}, \& {de Gregorio-Monsalvo}}]{Pinte:2016aa}
{Pinte}, C., {Dent}, W.~R.~F., {M{\'e}nard}, F., {et~al.} 2016, \apj, 816, 25

\bibitem[{{Pinte} {et~al.}(2022){Pinte}, {Teague}, {Flaherty}, {Hall},
  {Facchini}, \& {Casassus}}]{PinteTeague+:2022om}
{Pinte}, C., {Teague}, R., {Flaherty}, K., {et~al.} 2022, arXiv e-prints,
  arXiv:2203.09528

\bibitem[{{Raettig} {et~al.}(2021){Raettig}, {Lyra}, \&
  {Klahr}}]{RaettigLyra+:2021sb}
{Raettig}, N., {Lyra}, W., \& {Klahr}, H. 2021, \apj, 913, 92

\bibitem[{{Riols} \& {Lesur}(2018)}]{Riols:2018aa}
{Riols}, A., \& {Lesur}, G. 2018, \aap, 617, A117

\bibitem[{{Sano} {et~al.}(2000){Sano}, {Miyama}, {Umebayashi}, \&
  {Nakano}}]{SanoMiyama+:2000fo}
{Sano}, T., {Miyama}, S.~M., {Umebayashi}, T., \& {Nakano}, T. 2000, \apj, 543,
  486

\bibitem[{{Sekiya}(1998)}]{Sekiya:1998aa}
{Sekiya}, M. 1998, \icarus, 133, 298

\bibitem[{{Simon} {et~al.}(2013{\natexlab{a}}){Simon}, {Bai}, {Armitage},
  {Stone}, \& {Beckwith}}]{Simon+2013a}
{Simon}, J.~B., {Bai}, X.-N., {Armitage}, P.~J., {Stone}, J.~M., \& {Beckwith},
  K. 2013{\natexlab{a}}, \apj, 775, 73

\bibitem[{{Simon} {et~al.}(2013{\natexlab{b}}){Simon}, {Bai}, {Stone},
  {Armitage}, \& {Beckwith}}]{Simon+2013b}
{Simon}, J.~B., {Bai}, X.-N., {Stone}, J.~M., {Armitage}, P.~J., \& {Beckwith},
  K. 2013{\natexlab{b}}, \apj, 764, 66

\bibitem[{{Stoll} \& {Kley}(2014)}]{StollKley2014}
{Stoll}, M. H.~R., \& {Kley}, W. 2014, \aap, 572, A77

\bibitem[{{Stoll} \& {Kley}(2016)}]{StollKley:2016vp}
---. 2016, \aap, 594, A57

\bibitem[{{Stone} {et~al.}(2020){Stone}, {Tomida}, {White}, \&
  {Felker}}]{Stone:2020aa}
{Stone}, J.~M., {Tomida}, K., {White}, C.~J., \& {Felker}, K.~G. 2020, \apjs,
  249, 4

\bibitem[{{Takahashi} \& {Inutsuka}(2014)}]{Takahashi:2014wi}
{Takahashi}, S.~Z., \& {Inutsuka}, S.-i. 2014, \apj, 794, 55

\bibitem[{{Takeuchi} \& {Lin}(2002)}]{TakeuchiLin2002}
{Takeuchi}, T., \& {Lin}, D.~N.~C. 2002, \apj, 581, 1344

\bibitem[{{Tanaka} {et~al.}(2005){Tanaka}, {Himeno}, \&
  {Ida}}]{TanakaHimeno+:2005wc}
{Tanaka}, H., {Himeno}, Y., \& {Ida}, S. 2005, \apj, 625, 414

\bibitem[{{Teague} {et~al.}(2016){Teague}, {Guilloteau}, {Semenov}, {Henning},
  {Dutrey}, {Pi{\'e}tu}, {Birnstiel}, {Chapillon}, {Hollenbach}, \&
  {Gorti}}]{TeagueGuilloteau+:2016uj}
{Teague}, R., {Guilloteau}, S., {Semenov}, D., {et~al.} 2016, \aap, 592, A49

\bibitem[{{Teague} {et~al.}(2018){Teague}, {Henning}, {Guilloteau}, {Bergin},
  {Semenov}, {Dutrey}, {Flock}, {Gorti}, \&
  {Birnstiel}}]{TeagueHenning+:2018wx}
{Teague}, R., {Henning}, T., {Guilloteau}, S., {et~al.} 2018, \apj, 864, 133

\bibitem[{{Tominaga} {et~al.}(2018){Tominaga}, {Inutsuka}, \&
  {Takahashi}}]{Tominaga:2018th}
{Tominaga}, R.~T., {Inutsuka}, S.-i., \& {Takahashi}, S.~Z. 2018, \pasj, 70, 3

\bibitem[{{Tominaga} {et~al.}(2019){Tominaga}, {Takahashi}, \&
  {Inutsuka}}]{Tominaga:2019uu}
{Tominaga}, R.~T., {Takahashi}, S.~Z., \& {Inutsuka}, S.-i. 2019, \apj, 881, 53

\bibitem[{{Tominaga} {et~al.}(2020){Tominaga}, {Takahashi}, \&
  {Inutsuka}}]{Tominaga:2020wn}
---. 2020, \apj, 900, 182

\bibitem[{{Umurhan} {et~al.}(2020){Umurhan}, {Estrada}, \&
  {Cuzzi}}]{UmurhanEstrada+:2020yi}
{Umurhan}, O.~M., {Estrada}, P.~R., \& {Cuzzi}, J.~N. 2020, \apj, 895, 4

\bibitem[{{Urpin}(2003)}]{Urpin2003}
{Urpin}, V. 2003, \aap, 404, 397

\bibitem[{{Urpin} \& {Brandenburg}(1998)}]{UrpinBrandenburg1998}
{Urpin}, V., \& {Brandenburg}, A. 1998, \mnras, 294, 399

\bibitem[{{van der Marel} {et~al.}(2019){van der Marel}, {Dong}, {di
  Francesco}, {Williams}, \& {Tobin}}]{van-der-Marel:2019aa}
{van der Marel}, N., {Dong}, R., {di Francesco}, J., {Williams}, J.~P., \&
  {Tobin}, J. 2019, \apj, 872, 112

\bibitem[{{van Leer}(1974)}]{van-Leer:1974xu}
{van Leer}, B. 1974, Journal of Computational Physics, 14, 361

\bibitem[{{Villenave} {et~al.}(2022){Villenave}, {Stapelfeldt}, {Duch{\^e}ne},
  {M{\'e}nard}, {Lambrechts}, {Sierra}, {Flores}, {Dent}, {Wolff}, {Ribas},
  {Benisty}, {Cuello}, \& {Pinte}}]{VillenaveStapelfeldt+:2022pp}
{Villenave}, M., {Stapelfeldt}, K.~R., {Duch{\^e}ne}, G., {et~al.} 2022, \apj,
  930, 11

\bibitem[{{Weidenschilling}(1980)}]{Weidenschilling:1980xl}
{Weidenschilling}, S.~J. 1980, \icarus, 44, 172

\bibitem[{{Whipple}(1972)}]{Whipple:1972vv}
{Whipple}, F.~L. 1972, in From Plasma to Planet, ed. A.~{Elvius}, 211

\bibitem[{{Windmark} {et~al.}(2012){Windmark}, {Birnstiel}, {Ormel}, \&
  {Dullemond}}]{Windmark:2012aa}
{Windmark}, F., {Birnstiel}, T., {Ormel}, C.~W., \& {Dullemond}, C.~P. 2012,
  \aap, 544, L16

\bibitem[{{Yang} {et~al.}(2017){Yang}, {Johansen}, \& {Carrera}}]{Yang:2017aa}
{Yang}, C.~C., {Johansen}, A., \& {Carrera}, D. 2017, \aap, 606, A80

\bibitem[{{Youdin}(2011)}]{Youdin:2011aa}
{Youdin}, A.~N. 2011, \apj, 731, 99

\bibitem[{{Youdin} \& {Goodman}(2005)}]{Youdin:2005aa}
{Youdin}, A.~N., \& {Goodman}, J. 2005, \apj, 620, 459

\bibitem[{{Youdin} \& {Lithwick}(2007)}]{YoudinLithwick2007}
{Youdin}, A.~N., \& {Lithwick}, Y. 2007, \icarus, 192, 588

\bibitem[{{Youdin} \& {Shu}(2002)}]{Youdin:2002aa}
{Youdin}, A.~N., \& {Shu}, F.~H. 2002, \apj, 580, 494

\end{thebibliography}

\appendix

\section{Parameter choices and final saturated states for all runs}\label{appendix:parameter_sets}

    \begin{table}[h!]
    \tbl{Parameter choices of our simulations with $\beta$ cooling models and final saturated states for all runs.}{%
    \tabcolsep = 1.3mm
    \begin{tabular}{cccccccc}
    \hline
    Run Name & $a$ & $\beta_1$ & $b$ & $\Delta L_{\rm u}/H_{\rm g}$ & $\Delta L_{\rm s}/H_{\rm g}$ & Time {[$P_{\rm in}$]} & T/pT  \\
    \hline
    isothermal & - & - & - & - & - & 400 & - \\
    a270-b000 & 2.7 & 0.0 & - & 10.0 & 0.0 & 400 & T\\ 
    a270-b015 & 2.7 & 0.5 & 0.15 & 9.6 & 0.4 & 400 & T\\ 
    a270-b030 & 2.7 & 0.5 & 0.3 & 9.1 & 0.9 & 400 & T\\ 
    a270-b050 & 2.7 & 0.5 & 0.5 & 8.6 & 1.5 & 400 & T\\ 
    a270-b060 & 2.7 & 0.5 & 0.6 & 8.3 & 1.8 & 400 & T\\ 
    a270-b070 & 2.7 & 0.5 & 0.7 & 8.0 & 2.0 & 400 & T\\ 
    a270-b080 & 2.7 & 0.5 & 0.8 & 7.7 & 2.3 & 400 & T\\ 
    a270-b090 & 2.7 & 0.5 & 0.9 & 7.4 & 2.6 & 800 & pT\\
    a270-b100 & 2.7 & 0.5 & 1.0 & 7.1 & 2.9 & 800 & pT\\ 
    a270-b110 & 2.7 & 0.5 & 1.1 & 6.8 & 3.2 & 800 & pT\\
    a270-b120 & 2.7 & 0.5 & 1.2 & 6.5 & 3.5 & 800 & pT\\
    a270-b130 & 2.7 & 0.5 & 1.3 & 6.2 & 3.8 & 800 & pT\\
    a210-b000 & 2.1 & 0.0 & - & 7.8 & 0.0 & 400 & T\\ 
    a210-b015 & 2.1 & 0.5 & 0.15 & 7.3 & 0.4 & 400 & T\\ 
    a210-b030 & 2.1 & 0.5 & 0.3 & 6.9 & 0.9 & 400 & T\\ 
    a210-b050 & 2.1 & 0.5 & 0.5 & 6.3 & 1.5 & 400 & T\\ 
    a210-b060 & 2.1 & 0.5 & 0.6 & 6.0 & 1.8 & 400 & T\\ 
    a210-b070 & 2.1 & 0.5 & 0.7 & 5.8 & 2.1 & 400 & T\\ 
    a210-b080 & 2.1 & 0.5 & 0.8 & 5.5 & 2.3 & 800 & T\\ 
    a210-b090 & 2.1 & 0.5 & 0.9 & 5.2 & 2.6 & 800 & pT\\
    a210-b100 & 2.1 & 0.5 & 1.0 & 4.9 & 2.9 & 800 & pT\\ 
    a210-b110 & 2.1 & 0.5 & 1.1 & 4.6 & 3.2 & 800 & pT\\ 
    a150-b000 & 1.5 & 0.0 & - & 5.6 & 0.0 & 400 & T\\ 
    a150-b015 & 1.5 & 0.5 & 0.15 & 5.1 & 0.4 & 400 & T\\ 
    a150-b030 & 1.5 & 0.5 & 0.3 & 4.7 & 0.9 & 400 & T\\ 
    a150-b050 & 1.5 & 0.5 & 0.5 & 4.1 & 1.5 & 400 & T\\ 
    a150-b060 & 1.5 & 0.5 & 0.6 & 3.8 & 1.8 & 400 & T\\ 
    a150-b070 & 1.5 & 0.5 & 0.7 & 3.5 & 2.1 & 800 & T\\ 
    a150-b080 & 1.5 & 0.5 & 0.8 & 3.2 & 2.4 & 800 & T\\ 
    a150-b090 & 1.5 & 0.5 & 0.9 & 2.9 & 2.7 & 800 & T\\ 
    a100-b000 & 1.0 & 0.0 & - & 3.7 & 0.0 & 400 & T\\
    a100-b015 & 1.0 & 0.5 & 0.15 & 3.3 & 0.4 & 400 & T\\
    a100-b030 & 1.0 & 0.5 & 0.3 & 2.8 & 0.9 & 400 & T\\ 
    a100-b050 & 1.0 & 0.5 & 0.5 & 2.3 & 1.5 & 800 & T\\ 
    a100-b060 & 1.0 & 0.5 & 0.6 & 1.9 & 1.8 & 800 & T\\ 
    a100-b070 & 1.0 & 0.5 & 0.7 & 1.6 & 2.1 & 800 & T\\ 
    a100-b080 & 1.0 & 0.5 & 0.8 & 1.3 & 2.4 & 800 & T\\ 
    a075-b000 & 0.75 & 0.0 & - & 2.9 & 0.0 & 800 & T\\ 
    a075-b015 & 0.75 & 0.5 & 0.15 & 2.3 & 0.4 & 800 & T\\ 
    a075-b030 & 0.75 & 0.5 & 0.3 & 1.9 & 0.9 & 800 & T\\ 
    a075-b040 & 0.75 & 0.5 & 0.4 & 1.6 & 1.2 & 800 & T\\ 
    a075-b050 & 0.75 & 0.5 & 0.5 & 1.3 & 1.5 & 800 & T\\ 
    a050-b000 & 0.5 & 0.0 & - & 1.8 & 0.0 & 800 & T\\ 
    a050-b015 & 0.5 & 0.5 & 0.15 & 1.4 & 0.4 & 800 & T\\ 
    a050-b030 & 0.5 & 0.5 & 0.3 & 1.0 & 0.9 & 800 & T\\ 
    a025-b000 & 0.25 & 0.0 & - & 0.9 & 0.0 & 800 & T\\
    \hline
    \end{tabular}}\label{t:beta}
    \end{table}

    \begin{figure*}
        \begin{center}
        \includegraphics[width=\hsize,bb = 0 0 778 992]{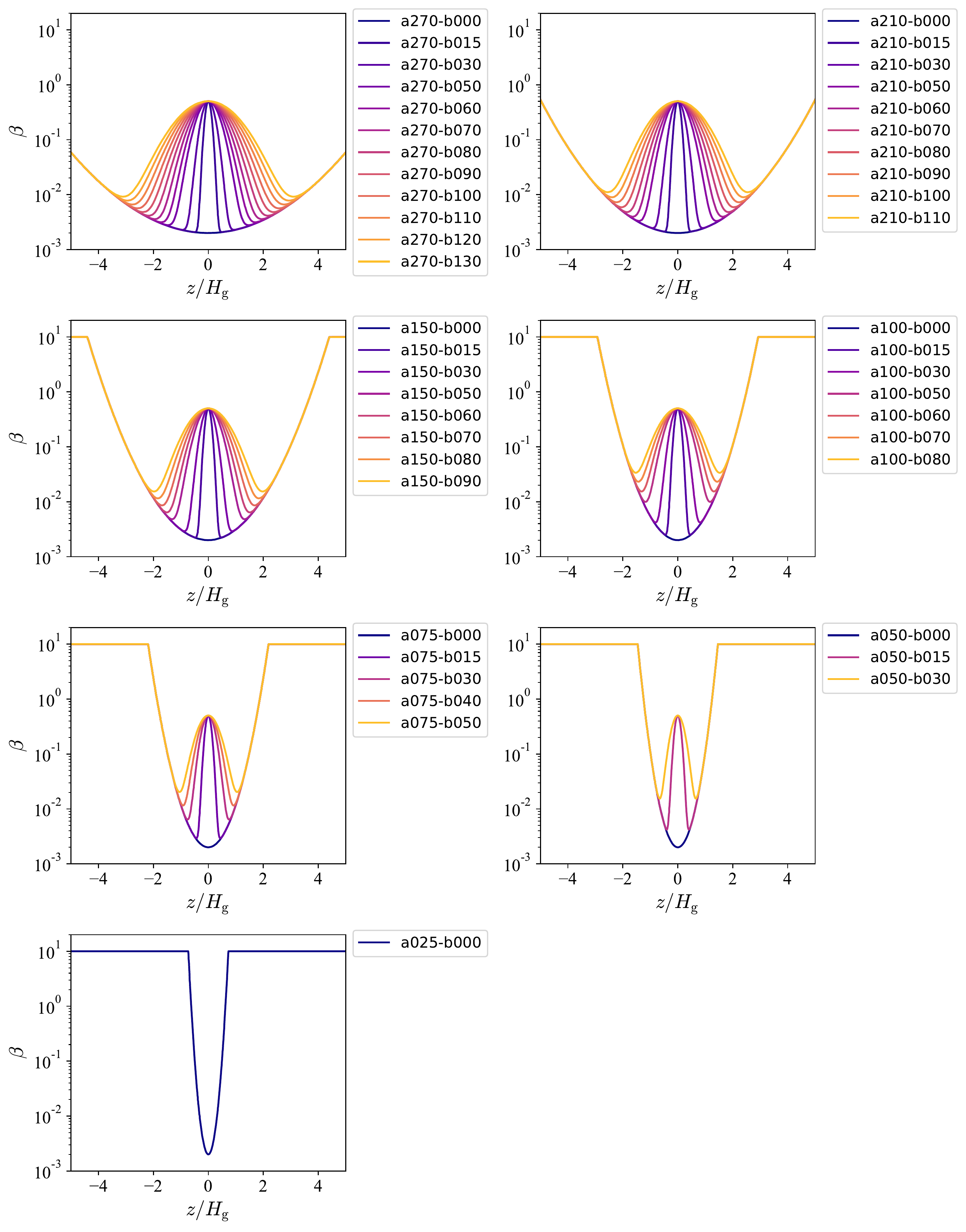}
        \end{center}
        \caption{Vertical profiles of dimensionless cooling time $\beta$ as a function $z/H_{\rm g}$ at $R=1.0$ for all runs presented in this study.}
        \label{fig:beta_model}
    \end{figure*}
    
    We perform 46 hydrodynamical simulations with different values of $a$, $\beta_1$, and $b$ that determine the vertical profile of the cooling time $\beta$ (equation \eqref{eq:beta_model}) and thicknesses of the unstable and midplane stable layers (equation \eqref{eq:global_criterion}).
    The values of $a$, $\beta_1$, $b$, $\Delta L_{\rm u}$, and $\Delta L_{\rm s}$ for all runs presented in this study are summarized in table \ref{t:beta}.
    Figure \ref{fig:beta_model} illustrates the vertical $\beta$ profiles for all runs.
    
    We stop a run at a different time for each run because the time, until the system reaches in quasi-steady state, differs from one run to another.
    The runtimes and final saturated states of all our simulations are also summarized in table \ref{t:beta}.
    
\section{Examples of turbulence time evolution and vertical structure for runs with T and pT states}\label{appendix:comparison}

\begin{figure}
    \begin{center}
    \includegraphics[width=80mm,bb = 0 0 354 615]{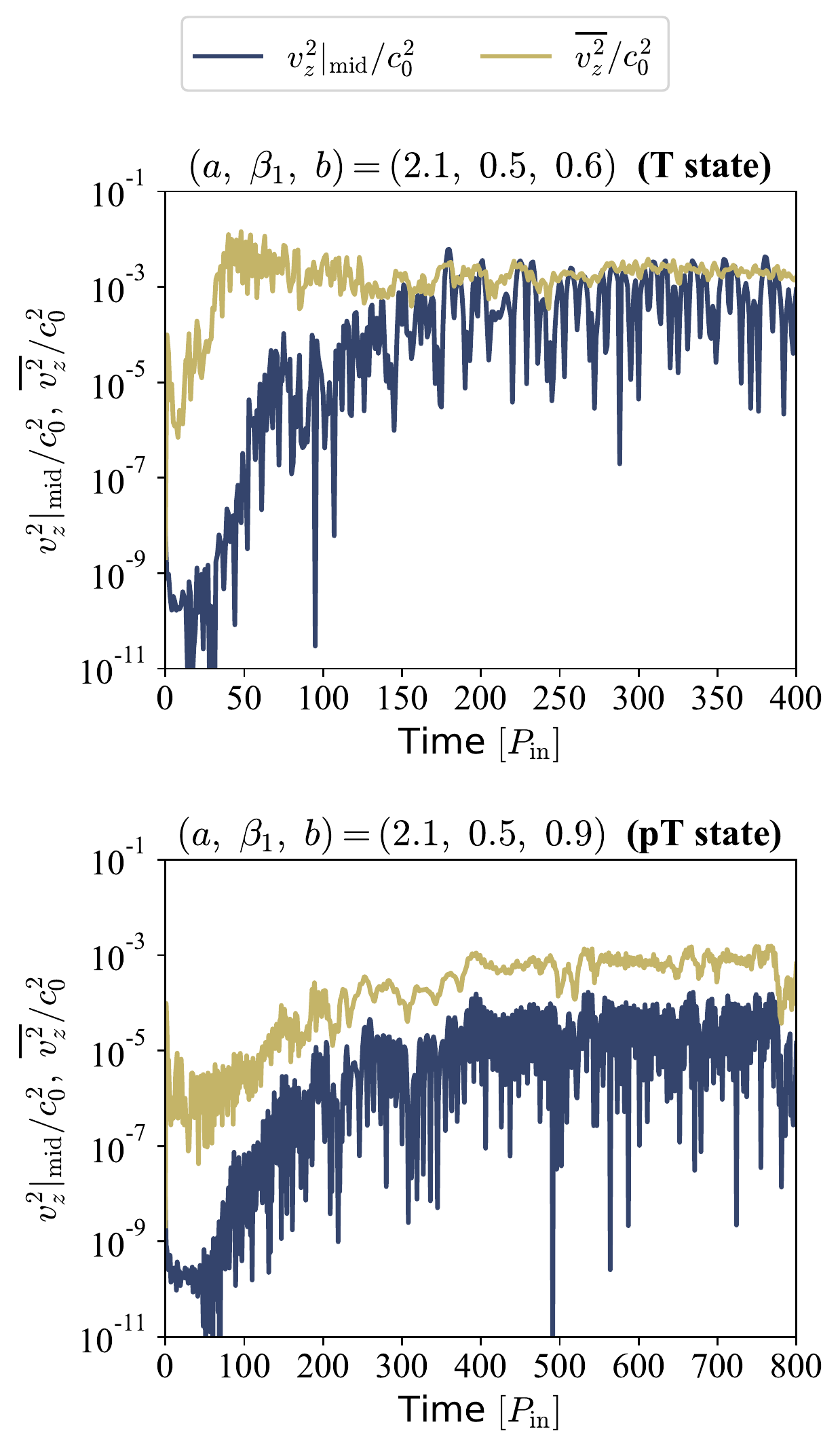}
    \end{center}
    \caption{Time evolution of the squared vertical velocity at the midplane $v_z^2|_{\rm mid}$ and $v_z^2$ averaged in vertical direction $\overline{v_z^2}$ for runs with $(a,~\beta_1,~b) = (2.1,~0.5,~0.6)$ (T state; upper panel) and $(a,~\beta_1,~b) = (2.1,~0.5,~0.9)$ (pT state; lower panel) at $R=1.0$.}
    \label{fig:vz2_TE}
\end{figure}

In figure \ref{fig:vz_colormap_tz}, we display the two-dimensional maps and vertical profile's time evolution of vertical velocity for runs with $(a,~\beta_1,~b) = (2.1,~0.5,~0.6)$ and $(2.1,~0.5,~0.9)$, which reach the T state and pT state, respectively.
These two runs also differ in their time evolution of $v_z^2|_{\rm mid}$ and $\overline{v_z^2}$.
Figure \ref{fig:vz2_TE} shows time evolution of $v_z^2|_{\rm mid}$ and $\overline{v_z^2}$ for the two runs presented in section \ref{subsec:overview}.
For $(a,~\beta_1,~b) = (2.1,~0.5,~0.6)$ (T state), the vertical flows develop at $\lesssim 50 P_{\rm in}$, corresponding to the growth rate $\Gamma_{\rm VSI}$ being $\sim 10^{-2}\Omega_{\rm K}$.
On the other hand, for $(a,~\beta_1,~b) = (2.1,~0.5,~0.9)$ (pT state), the vertical gas motion grows at $\lesssim 300 P_{\rm in}$ with $\Gamma_{\rm VSI} \sim 10^{-3} \Omega_{\rm K}$.
The values of $v_z^2|_{\rm mid}$ and $\overline{v_z^2}$ are steady over $t = 250$--$400$ orbits and $t = 650$--$800$ orbits for runs with the T state and pT state, respectively, which are consistent with the vertical velocity's time evolution shown in figure \ref{fig:vz_colormap_tz}.

\begin{figure}
    \begin{center}
    \includegraphics[width=80mm,bb = 0 0 361 594]{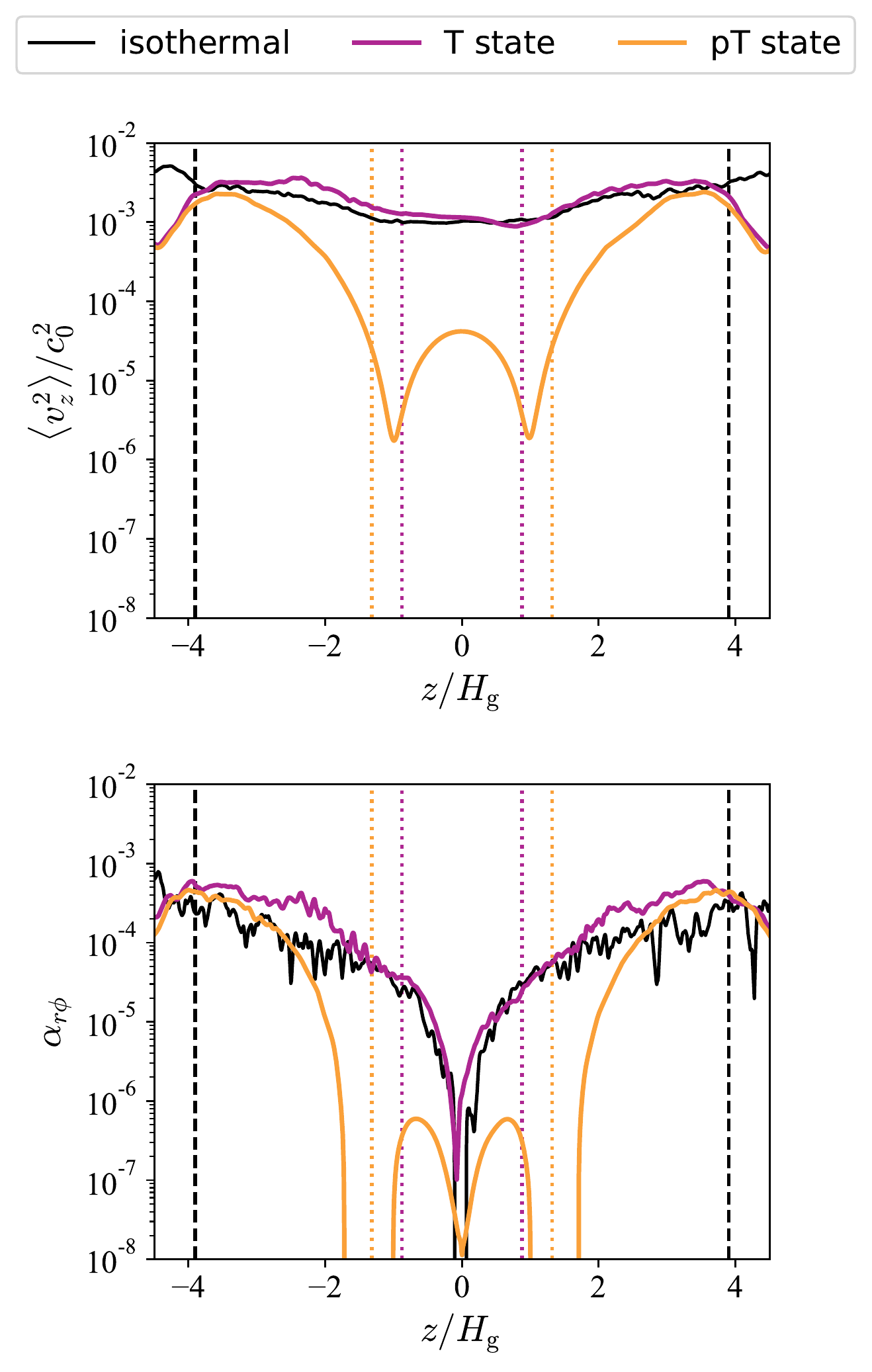}
    \end{center}
    \caption{Vertical profile of time mean squared vertical velocity $\langle v_z^2 \rangle$ (upper panel) and Reynolds stress $\alpha_{r\phi}$ (lower panel) for runs with the isothermal state, $(a,~\beta_1,~b) = (2.1,~0.5,~0.6)$ (T state), and $(a,~\beta_1,~b) = (2.1,~0.5,~0.9)$ (pT state) at $R=1.0$. The dashed dotted lines represent the height of the unstable layer's upper boundaries $z = \pm z_{\rm u}$, which is the same for these two runs. The dotted lines represent the height of the midplane stable layer's uuper boundaries $z = \pm z_{\rm s}$ for each run.}
    \label{fig:vz2_alpha_R1}
\end{figure}

These two runs reach different quasi-steady states, resulting in different turbulence profiles (see figures \ref{fig:vz_colormap_Rz} and \ref{fig:vz_colormap_tz}).
To evaluate this quantitatively, we plot in figure \ref{fig:vz2_alpha_R1} vertical profiles of $\langle v_z^2 \rangle$ and $\alpha_{r\phi}$ for these two cases.
In both cases, $\langle v_z^2 \rangle$ and $\alpha_{r\phi}$ have a vertical profile with peaks near the unstable layer's upper boundary at $z = \pm z_{\rm u}$.
The run with $(a,~\beta_1,~b) = (2.1,~0.5,~0.6)$ (T state) has almost the same turbulent structure as the run of the isothermal state, which is the fundamental ideal state for the VSI.
In contrast, for $(a,~\beta_1,~b) = (2.1,~0.5,~0.9)$ (pT state), $\langle v_z^2 \rangle$ as well as $\alpha_{r\phi}$ decreases sharply near the stable midplane layer's upper boundary at $z = \pm z_{\rm s}$.
This vertical distribution of $\langle v_z^2 \rangle$ yields the difference of more than an order of magnitude between $\langle v_z^2 \rangle|_{\rm mid}$ and $\langle \overline{v_z^2} \rangle$ (see figure \ref{fig:vz2_mid_intz_R1}).
The increase in $\langle v_z^2 \rangle$ around the midplane may be related to the linear VSI local criterion in appendix \ref{appendix:local_criterion}.
Additionally, figure \ref{fig:vz2_alpha_R1} indicates that $\alpha_{r\phi}$ is an order-of-magnitude smaller than $\langle v_z^2 \rangle$ for both the T and pT states.
This is consistent with the correlation between $\overline{\alpha_{r\phi}}$ and $\langle \overline{v_z^2}\rangle$ shown in figure \ref{fig:vz2_intz_alpha}.

\section{Local criterion of linear VSI}\label{appendix:local_criterion}

So far we have used the global criterion of the linear VSI in equation \eqref{eq:global_criterion} to determine the thicknesses of the unstable and stable layers.
However, the slight rise of $\langle v_z^2 \rangle$ near the midplane of the run with $(a,~\beta_1,~b) = (2.1,~0.5,~0.9)$ (pT state) shown in the upper panel of figure \ref{fig:vz2_alpha_R1} may be due to an unstable midplane layer with a slightly thin determined by a local criterion of the VSI.

The onset of the VSI requires reducing buoyancy by rapid disk cooling.
When this effect is compared with the vertical shear that drives the VSI, the local criterion is given by \citep{Urpin2003,LinYoudin2015}
\begin{equation}\label{eq:local_criterion}
    \beta \leq \beta_{\rm lc} =  \frac{|\pd_z(R\Omega)|}{N_z^2}\Omega_{\rm K}.
\end{equation}
Here, $\beta_{\rm lc}$ is the local critical dimensionless cooling time and $N_z$ is the Brunt--V\"{a}is\"{a}l\"{a} frequency defined by
\begin{equation}
    N_z^2 \equiv -\frac{1}{\rho_{\rm g} C_P}\cdot\frac{\partial P}{\partial z}\cdot\frac{\partial s}{\partial z},
\end{equation}
where $C_P$ and $s$ are the specific heat at constant pressure and the specific entropy, respectively.
The specific entropy is given by $s = C_V\ln{\paren{P/\rho_{\rm g}^\gamma}}$, where $C_V$ is the specific heat at constant volume and $\gamma$ is the heat capacity ratio.
The local criterion can provide the global criterion in equation \eqref{eq:global_criterion} at $z = \gamma H_{\rm g}/2$ \citep{LinYoudin2015}.

The presence of the thin unstable midplane layer can be confirmed by looking at the vertical dependence of the local criterion.
Using the ideal gas law $P = k_{\rm B}\rho_{\rm g}T/m_{\rm g}$, where $k_{\rm B}$ is the Boltzmann constant and $m_{\rm g}$ is the mean molecular mass of the gas, and assuming vertically isothermal and vertical hydrostatic equilibrium (equation \eqref{eq:initial_gas_density}), we get $N_z^2 \propto (\pd_z \ln\rho_{\rm g})^2 \propto z^2$.
With this assumption and $\pd_z (R\Omega) \propto z$ (see equation \eqref{eq:vertical_shear}), the local dimensionless critical cooling time can be analytically performed, resulting in $\beta_{\rm lc} \propto |z|^{-1}$.
Therefore, in regions where $z$ is sufficiently small (but not zero), $\beta_{\rm lc}$ increases, and thus the local criterion in equation \eqref{eq:local_criterion} is satisfied for any finite $\beta$.
As a result, the thin VSI-active layer determined by the local criterion always exists near the midplane\footnote{For constant $\beta = 1$ in vertical direction, a thickness of the midplane unstable layer is $\sim 0.1H_{\rm g}$.}, although the VSI does not operate the midplane ($z = 0$) where the vertical shear vanishes.

This thin unstable layer near the midplane may be weakly turbulent by the VSI.
This may produce the slight vertical shading visible near the midplane in the right panel of figure \ref{fig:vz_colormap_tz} and the slight increase of $\langle v_z^2 \rangle$ near the midplane in the upper panel of figure \ref{fig:vz2_alpha_R1}.
Therefore, to discuss exactly where the VSI operates turbulence, it may be necessary to apply the local criterion as well as the global criterion.
 
\end{document}